\documentclass[%
showkeys,
preprint,
showpacs,preprintnumbers,
 amsmath,amssymb,
 aps,
]{revtex4-1}

\usepackage{graphicx}
\usepackage{dcolumn}
\usepackage{bm}
\usepackage{hyperref}

\usepackage{xcolor}

\usepackage[titletoc,title]{appendix}

\begin{document}

\preprint{PREPRINT}

\title{Mechanical characterization of disordered and anisotropic cellular monolayers}

\author{Alexander Nestor-Bergmann}
\affiliation{Department of Physiology, Development \& Neuroscience, University of Cambridge, Downing Street, Cambridge CB2 3DY UK
}%

\author{Emma Johns}%
\affiliation{Wellcome Trust Centre for Cell-Matrix Research, School of Medical Sciences, Faculty of Biology, Medicine \& Health, \\ Manchester Academic Health Science Centre, University of Manchester, Oxford Road, Manchester M13 9PT UK
}%
 
\author{Sarah Woolner}%
\affiliation{Wellcome Trust Centre for Cell-Matrix Research, School of Medical Sciences, Faculty of Biology, Medicine \& Health, \\  Manchester Academic Health Science Centre, University of Manchester, Oxford Road, Manchester M13 9PT UK
}%


\author{Oliver E.  Jensen}
\email{oliver.jensen@manchester.ac.uk}
\affiliation{School of Mathematics, University of Manchester, Manchester M13 9PL, UK}%

\date{\today}

\begin{abstract}
We consider a cellular monolayer, described using a vetex-based model, for which cells form a spatially disordered array of convex polygons that tile the plane.  Equilibrium cell configurations are assumed to minimize a global energy defined in terms of cell areas and perimeters; energy is dissipated via dynamic area and length changes, as well as cell neighbour exchanges.  The model captures our observations of an epithelium from a \textit{Xenopus} embryo showing that uniaxial stretching induces spatial ordering, with cells under net tension (compression) tending to align with (against) the direction of stretch, but with the stress remaining heterogeneous at the single-cell level.  We use the vertex model to derive the linearized relation between tissue-level stress, strain and strain-rate about a deformed base state, which can be used to characterize the tissue's anisotropic mechanical properties;  expressions for viscoelastic tissue moduli are given as direct sums over cells.  When the base state is isotropic, the model predicts that tissue properties can be tuned to a regime with high elastic shear resistance but low resistance to area changes, or \textit{vice versa}.
\end{abstract}

\maketitle


\section{Introduction}

Epithelial tissues have significant roles in embryonic development, tissue homeostasis and disease development \citep{guillot2013mechanics}.  Recent work has revealed that many critical functions in biological tissues are dependent on the accurate organisation of the shapes and packing geometry of the constituent cells \citep{lecuit2007cell}.  Disturbances in this organisation have been associated with problems during embryonic development and diseases in adult life \citep{carney1984mechanisms, jiao2011spatial}.  Furthermore, there is evidence that mechanical forces may directly trigger biochemical responses that regulate morphogenetic processes \citep{schluck2013mechanical, heisenberg2013forces}.   However, due to difficulties in quantifying stresses in tissues, the mechanisms by which tissue behaviour emerges from these multiscale feedback processes remain poorly understood.

Continuum descriptions can provide useful insights to tissue-level behaviour.  For example, elastic-viscoplastic continuum models can capture the solid- and liquid-like response of tissues to small and large deformations over differing timescales \citep{preziosi2010elasto, weickenmeier2014elastic}.  However, they are in many cases not built from explicit physical description of cells. Furthermore,  the interactions of multiple cells can lead to rich emergent behaviour at the tissue scale, such as yielding and remodelling, that is not easily accessed through conventional continuum frameworks  \citep{pathmanathan2009computational, ishihara2017cells}.  

Discrete vertex-based models of epithelia have been a useful tool in linking mechanics to tissue morphology \citep{farhadifar2007, nagai2001dynamic, fletcher2014vertex, alt2017, fletcher2017}.  These models have more recently been developed to characterize the mechanical properties of tissues \citep{murisic2015, merzouki2016} and to infer local and global stresses \citep{brodland2014, sugimura2013, nestorbergmann2017}.  This work has predicted interesting long-range mesoscopic mechanical patterning arising purely from the mechanical properties and short-range mechanical interactions of cells within the tissue, which are not seen in traditional continuum descriptions \citep{nestorbergmann2017, yang2017correlating}.  Relationships between discrete models and traditional continuum approaches have been found for spatially periodic cell networks \citep{murisic2015, staple2010mechanics}, while equivalent relationships for disordered tissues have only been partially established for isotropic disordered cell networks \citep{nestorbergmann2017} or for analogous physical systems such as two-dimensional (2D) dry foams \cite{kruyt2007}.

Many mechanical models of biological tissues assume that the material is isotropic.  However, recent observations in the \textit{Drosophila melanogaster} embryo \citep{gao2016embryo} have provided evidence that  biological tissues may exhibit orientational, as well as positional, structure.  Likewise, models of 3D foams have explored how orientational structure can be introduced through uniaxial deformations \citep{evans2017geometric}.  The deformation induces a net stress in the material, leading to a series of irreversible deformations (such as neighbour exchanges).  Experimental observations of epithelial tissues have revealed similar patterns of orientational order following stretching \citep{sugimura2013, nestorbergmann2017_bio, harris2012}.

To explore how deformation induces anisotropy, in this paper we use a variant of a well-studied vertex-based model to quantify the mechanical behaviour of a disordered cellular monolayer under an external load.  We ignore cell division or motility but take account of dissipation arising from changes in cell area and perimeter, motivated by observations of damping in the cytosol of cells in the \textit{Drosophila} embryo on a timescale of minutes \cite{doubrovinski2017}, evidence of shorter but distinct stress-relaxation timescales in the cytosol and cortex \cite{bambardekar2015}, sub-cellular observations of dissipation at cell contacts \cite{clement2017} {\color{red}and evidence of viscoelastic stress-relaxation in a freely suspended cultured monolayer \citep{harris2012}}.  We provide observations of stretched \textit{Xenopus} embryonic epithelium demonstrating that a uniaxial stretch enforces order in the tissue, in which cells under net tension tend to align their principal axis of stress with the stretch direction and those under compression align perpendicularly. This behaviour is captured in simulations, which are further used to quantify the tissue's anisotropy using the deviatoric (shear) component of the global stress.  We then derive a linearized stress/strain/strain-rate relationship characterizing the perturbation stress of a pre-stressed tissue subjected to a small homogeneous deformation.  This allows viscoelastic moduli to be computed for an anisotropic disordered cellular monolayer.  Finally, we show that the mechanical parameters of an isotropic tissue can be tuned to elicit high shear resistance but low resistance to area changes, or \textit{vice versa}.  

\section{Methods}

We use a modification of a popular vertex-based model to describe a planar epithelium \citep{farhadifar2007, fletcher2014vertex, bi2015}, using the notational framework presented in \cite{nestorbergmann2017}.  Details of our experimental protocol follow a summary of the model.

\subsection{The vertex-based model}

\begin{figure}
	\centering
	\includegraphics[width=0.9\textwidth]{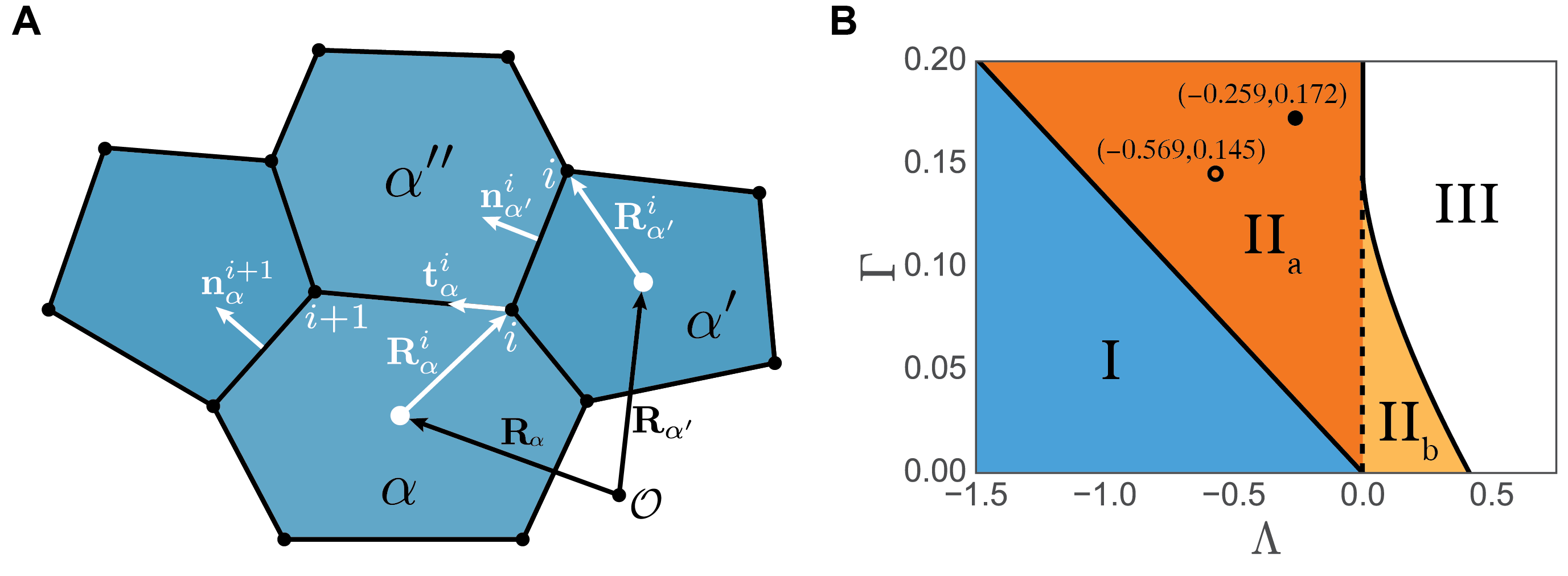}
	\caption{\textbf{A.}  Vertex model representation of tissue geometry.  The centroid of cell $\alpha$ is located at $\mathbf{R}_{\alpha}$ relative to the fixed origin, $\mathcal{O}$.  The position of cell vertices, $\mathbf{R}^{i}_{\alpha}$, are given relative to the centroid of a cell. Each vertex has three vectors, $\mathbf{R}^{i}_{\alpha}, \mathbf{R}^{i^{\prime}}_{\alpha^{\prime}}, \mathbf{R}^{i^{\prime\prime}}_{\alpha^{\prime\prime}}$, pointing to the same vertex from cells, $\alpha, \alpha^{\prime}, \alpha^{\prime\prime}$. Cell properties, such as area and tangents along edges, are also defined relative to the cell centroid.   \textbf{B.} $(\Lambda, \Gamma)$-parameter space, showing regimes in which tissues exhibit distinct behaviour. Following \citep{farhadifar2007}, region I is a `soft' network with no shear resistance; the network becomes solid-like in region II.  For hexagons, $P_6^{\text{eff}}=0$ (see (\ref{eq:p6})) has a single positive root in region II$_a$ and two positive roots in region $\text{II}_{b}$.  The network collapses in Region III.  Circular markers indicate locations of parameters used for simulations in Figure~\ref{fig:stress_vs_stretch}B below. }
    \label{fig:geometry}
\end{figure}

The epithelial monolayer, $\mathcal{M}$, is represented as a spatially disordered planar network of $N_{v}$ vertices, labelled $j=1,\dots, N_{v}$, connected by straight edges and bounding $N_{c}$ convex polygonal cells, labelled $\alpha=1,\dots,N_{c}$.   The cells are assumed to have identical mechanical properties described in terms of a preferred area $A_0^*$, a preferred perimeter $L_0^*$, a bulk stiffness $K^*$, a cortical stiffness $\Gamma^*$, a bulk viscosity $\gamma^*$ and a cortical viscosity $\mu^*$.   Scaling all distances on $\sqrt{A_0^*}$, the vector from the coordinate origin to vertex $j$ is given by ${\mathbf{R}}^{j}$.  Each vertex is shared by three cells and edges are shared by two cells (excluding cells at the boundary of $\mathcal{M}$). Quantities specific to cell $\alpha$ are labelled by a Greek subscript and defined relative to its centroid ${\mathbf{R}}_\alpha$ (Figure~\ref{fig:geometry}A). 

Cell $\alpha$ has $Z_\alpha$ vertices labelled anticlockwise by $i = 0,1,2,\dots, Z_\alpha-1$ relative to ${\mathbf{R}}_\alpha$.  We define $\mathbf{R}_\alpha^{i}$ as the vector from the cell centroid to vertex $i$, such that $\sum_{i=0}^{Z_\alpha-1} {\mathbf{R}}_\alpha^i=\mathbf{0}$.  Anticlockwise tangents are defined by ${\mathbf{t}}_\alpha^{i} = {\mathbf{R}}_\alpha^{i+1} - {\mathbf{R}}_\alpha^{i}$ (taking $i+1$ modulo $Z_\alpha$), unit vectors along a cell edge by $\hat{\mathbf{t}}_\alpha^i$ and outward normals to edges by ${\mathbf{n}}_\alpha^i={\mathbf{t}}_\alpha^{i} \times \hat{\mathbf{z}}$ (where $\hat{\mathbf{z}}$ is a unit vector pointing out of the plane).  The length, $l_\alpha^{i}$, of an edge belonging to cell $\alpha$ between vertices $i$ and $i+1$, the cell perimeter, $L_\alpha$, and the cell area, $A_\alpha$, are given by
\begin{equation}
\label{eq:1}
    \begin{gathered}
	l_\alpha^{i} = \left({\mathbf{t}}_\alpha^i\cdot {\mathbf{t}}_\alpha^i \right)^{1/2},\qquad
		{L}_\alpha=\sum_{i=0}^{Z_\alpha-1} {l}_\alpha^{i}, \qquad
        {A}_\alpha=\sum_{i=0}^{Z_\alpha-1} \tfrac{1}{2} \hat{\mathbf{z}} \cdot ({\mathbf{R}}_\alpha^i\times{\mathbf{R}}_\alpha^{i+1}).
    \end{gathered}
\end{equation}
We note for later reference that $\nabla_{\mathbf{R}_\alpha^i}A_\alpha=\mathbf{p}_\alpha^i\equiv \tfrac{1}{2}(\mathbf{n}_\alpha^i+\mathbf{n}_\alpha^{i-1})$ and $\nabla_{\mathbf{R}_\alpha^i}L_\alpha=- \mathbf{q}_\alpha^i$ where $\mathbf{q}_\alpha^i \equiv \hat{\mathbf{t}}_\alpha^i-\hat{\mathbf{t}}_\alpha^{i-1}$; furthermore  \cite{nestorbergmann2017}, 
\begin{equation}
\sum_{i=0}^{Z_\alpha-1} \mathbf{R}_\alpha^i \otimes \mathbf{p}_\alpha^i=A_\alpha\mathsf{I},\quad 
\sum_{i=0}^{Z_\alpha-1} \mathbf{R}_\alpha^i \otimes \mathbf{q}_\alpha^i=-L_\alpha\mathsf{Q}_\alpha
\equiv -\sum_{i=0}^{Z_\alpha-1} l_\alpha^i \hat{\mathbf{t}}_\alpha^i \otimes \hat{\mathbf{t}}_\alpha^i.
\label{eq:qa}
\end{equation}
$\mathsf{Q}_\alpha$ is a symmetric tensor characterizing the shape of cell $\alpha$, satisfying $\mathrm{Tr}(\mathsf{Q}_\alpha)=1$.

The dimensionless mechanical energy of an individual cell, $U_\alpha$ (scaled on $K^*A_0^{*2}$), is assumed to be 
\begin{equation}
    \label{eq:final_nondim}
    U_\alpha = \tfrac{1}{2}\left( A_\alpha - 1 \right)^{2} + \tfrac{1}{2} \Gamma \left( L_\alpha -L_0 
    \right)^{2}.
\end{equation}
Here the dimensionless parameter $\Gamma=\Gamma^*/(K^*A_0^*)$ represents the stiffness of the cell's cortex relative to its bulk; the preferred cell perimeter, $L_{0} =L_0^*/\sqrt{A_0^*}$ is often expressed in terms of a second dimensionless parameter $\Lambda=-2\Gamma L_0$.  Major features of the $(\Lambda,\Gamma)$-parameter space are shown in Figure~\ref{fig:geometry}B.   A rigidity transition characteristic of a glassy material takes place along $L_0 = \mu_6$ for regular hexagons \citep{farhadifar2007, Staple:2010}, where $\mu_Z=2(Z \tan(\pi/Z))^{1/2}$ (a regular $Z$-gon has an exact perimeter area relationship, $L=\mu_Z \sqrt{A}$).  For a disordered monolayer, the transition occurs along $L_0 \approx \mu_5$ \citep{bi2015}.  The transition between the fluid regime (I) and the solid regime (II) is indicated in Figure~\ref{fig:geometry}B.  We avoid region III below, where the network collapses.

We label derivatives $\partial U_\alpha/\partial A_\alpha$ and $\partial U_\alpha/\partial L_\alpha$ of (\ref{eq:final_nondim}) as a pressure and a tension respectively:
\begin{equation}
\label{eq:pt}
 P_\alpha \equiv A_\alpha-1,\quad T_\alpha\equiv \Gamma (L_\alpha-L_0).  
 \end{equation}
 Setting $\mathbf{f}_\alpha^{i} \equiv \nabla_{\mathbf{R}_\alpha^{i}} U_\alpha=P_\alpha \mathbf{p}_\alpha^i-T_\alpha \mathbf{q}_\alpha^i$, we can then interpret $-\mathbf{f}_\alpha^i$ as the elastic restoring force generated by cell $\alpha$ when vertex $i$ undergoes a small displacement.  

In a departure from many previous models we introduce the cell's (dimensionless) energy dissipation rate as
\begin{equation}
\Phi_\alpha= \gamma \dot{A}_\alpha^2+\mu \dot{L}_\alpha^2\equiv \gamma \sum_{i,j} \dot{\mathbf{R}}_\alpha^{i} \cdot \mathsf{A}_\alpha^{ij} \cdot \dot{\mathbf{R}}_\alpha^j + \mu \sum_{i,j}\dot{\mathbf{R}}_\alpha^{i} \cdot \mathsf{L}_\alpha^{ij} \cdot \dot{\mathbf{R}}_\alpha^j,
\label{eq:diss}
\end{equation}
where a dot {\color{red}above a variable} denotes a time derivative.  This accounts for viscous dissipation associated with shape changes of individual cells, {\color{red}and is expressed in terms of the two geometric variables characterizing cell shape, $A_\alpha$ and $L_\alpha$, that appear in the strain energy $U_\alpha$; this model does not describe frictional interaction with a substrate}.  It follows from (\ref{eq:1}) that $\dot{A}_\alpha=\sum_i \mathbf{p}_\alpha^i \cdot \dot{\mathbf{R}}_\alpha^i$, $\dot{L}_\alpha=-\sum_i \mathbf{q}_\alpha^i \cdot \dot{\mathbf{R}}_\alpha^i$, so that $\mathsf{A}_\alpha^{ij}\equiv\mathbf{p}_\alpha^i \otimes \mathbf{p}^j_\alpha$ and $\mathsf{L}_\alpha^{ij}\equiv\mathbf{q}_\alpha^i \otimes \mathbf{q}^j_\alpha$.  The parameters $\gamma$ and $\mu$ can be related to their dimensional counterparts via $\gamma=\gamma^*/K^*T^*$ and $\mu=\mu^*L_0^{*2}/K^* T^* A_0^*$ through a choice of timescale $T^*$ that we do not specify immediately.  It follows from (\ref{eq:diss}) that $\Phi_\alpha=\tfrac{1}{2}\sum_i \dot{\mathbf{R}}_\alpha^i \cdot \nabla_{\dot{\mathbf{R}}_\alpha^i} \Phi_\alpha$ and $\tfrac{1}{2}\nabla_{\dot{\mathbf{R}}_\alpha^i}\Phi_\alpha=\gamma \dot{A}_\alpha \mathbf{p}_\alpha^i - \mu \dot{L}_\alpha \mathbf{q}_\alpha^i$.  Following Fozard et al. \cite{fozard2009}, who treated the analogous 1D  problem, we minimize the total dissipation rate across the monolayer, $\Phi=\sum_\alpha \Phi_\alpha$, subject to a constraint ensuring the dissipation of total mechanical energy $U=\sum_\alpha U_\alpha$ through $\Phi$,
\begin{equation}
\dot{U}\equiv \sum_{\alpha=1}^{N_c} \sum_{i=1}^{Z_\alpha-1} \dot{\mathbf{R}}_\alpha^i \cdot \mathbf{f}_\alpha^i = -\Phi.
\label{eq:cons}
\end{equation}
This is achieved by minimizing the Lagrangian $\mathcal{L}=\Phi+\zeta(\dot{U}+\Phi)$ for some Lagrange multiplier $\zeta$.  The first variation of $\mathcal{L}$ with respect to the velocity of each vertex in $\mathcal{M}$ must vanish, \hbox{i.e.}
\begin{equation}
\nabla_{\dot{\mathbf{R}}^j} \mathcal{L}=\sum_{\{\mathbf{R}_\alpha^i=\mathbf{R}^j\}} \left[ (1+\zeta) \nabla_{\dot{\mathbf{R}}_\alpha^i} \Phi_\alpha+\zeta \mathbf{f}_\alpha^i\right] =\mathbf{0}, \quad (j=1,2,\dots,N_v)
\label{eq:cons2}
\end{equation}
where for each $j$ the sum is over the three cells adjacent to $\mathbf{R}^j$; likewise $\mathcal{L}_\zeta=0$ yields (\ref{eq:cons}).  Acting on (\ref{eq:cons2}) with 
$\sum_j \dot{\mathbf{R}}^j \cdot$ yields $(1+\zeta) 2\Phi + \zeta \dot{U}=0$, which with (\ref{eq:cons}) implies $\zeta=-2$ and hence (\ref{eq:cons2}) gives the net force balance on each vertex as $2\mathbf{F}^j=\mathbf{0}$ ($j=1,2,\dots,N_v$) where
\begin{equation}
\mathbf{F}^j\equiv \sum_{\{\mathbf{R}_\alpha^i=\mathbf{R}^j\}} \mathbf{F}_\alpha^i, \quad\mathrm{where}\quad 
\mathbf{F}_\alpha^i= - (P_\alpha+\gamma \dot{A}_\alpha) \mathbf{p}_\alpha^i + (T_\alpha+\mu \dot{L}_\alpha ) \mathbf{q}_\alpha^i .
\label{eq:force}
\end{equation}
$\mathbf{F}_\alpha^i$ can be interpreted as the viscoelastic restoring force due to cell $\alpha$ alone following a small displacement of its $i$th vertex.

It is computationally convenient (particularly when modelling viscous interaction with a substrate) simply to impose a drag on each vertex, leading to an explicit set of ODEs (of the form $\dot{\mathbf{R}}_j\propto -\sum_{\{\mathbf{R}_\alpha^i=\mathbf{R}_j\}}\mathbf{f}_\alpha^i$) that can be used to step the system forward in time; in contrast, (\ref{eq:force}) couples time derivatives in a more complex manner, falling into a class of models reviewed in \cite{alt2017}.  The tradeoff is a formulation that combines elastic responses to area and perimeter changes (via (\ref{eq:pt})) with their natural viscous counterparts, leading to an expression for the cell stress tensor $\boldsymbol{\sigma}_\alpha=(1/A_\alpha)\sum_i \mathbf{R}_\alpha^i \otimes \mathbf{F}_\alpha^i$ of the form (using (\ref{eq:qa}) and (\ref{eq:force}))
\begin{equation}
\boldsymbol{\sigma}_\alpha=-(P_\alpha+\gamma \dot{A}_\alpha) \mathsf{I}-(T_\alpha+\mu \dot{L}_\alpha ) \frac{L_\alpha}{A_\alpha} \mathsf{Q}_\alpha.
\label{eq:stres1}
\end{equation}
This retains the property that the principal axes of cell stress and cell shape (as defined by a shape tensor based on vertex locations) are aligned \citep{nestorbergmann2017}.

\subsection{Tissue-level stress}

The stress tensor for a single cell (\ref{eq:stres1}) may be rewritten as
\begin{equation}
    \label{eq:full_stress}
    \boldsymbol{\sigma}_\alpha = -P^{\text{eff}}_\alpha\mathsf{I} + (T_\alpha +\mu \dot{L}_\alpha)(L_\alpha/A_\alpha) {\mathsf{J}}_\alpha,
\end{equation}
where $\mathsf{J}_\alpha=\tfrac{1}{2}\mathsf{I}-\mathsf{Q}_\alpha$ is deviatoric, satisfying $\mathrm{Tr}(\mathsf{J}_\alpha) = 0$.  The isotropic component of the stress is given in terms of the effective cell pressure, defined as
\begin{equation}
P^{\text{eff}}_\alpha = P_\alpha +\gamma \dot{A}_\alpha + \tfrac{1}{2} (T_\alpha +\mu \dot{L}_\alpha) (L_\alpha / A_\alpha).
\label{eq:peff}
\end{equation}
Positive (negative) values of $P^{\text{eff}}_\alpha$ indicate that the cell is under net tension (compression). 

Assuming the monolayer forms a simply connected region of tissue, the tissue-level stress, $\boldsymbol{\sigma}^\mathcal{M}$, satisfies \cite{nestorbergmann2017}
\begin{equation}
	\label{eq:global_prestress}
	A^{\mathcal{M}} \boldsymbol{\sigma}^{\mathcal{M}} = \sum_\alpha A_\alpha \boldsymbol{\sigma}_\alpha,
\end{equation}
where the sum is over all cells in $\mathcal{M}$ and $A^{\mathcal{M}}=\sum_\alpha A_\alpha$.  Correspondingly, the isotropic component of tissue level stress is expressed in terms of {\color{red}the effective tissue pressure}
\begin{equation}
	 \label{eq:glob_pre}
	\overline{P^{\mathrm{eff}}} \equiv \frac{1}{A^{\mathcal{M}}} \sum_\alpha A_\alpha P^{\mathrm{eff}}_\alpha =  -\tfrac{1}{2}\mathrm{Tr}( \boldsymbol{\sigma}^{\mathcal{M}} ).
\end{equation}
For a monolayer under isotropic external loading, the deviatoric component of the global stress must vanish.  Thus, once in equilibrium, the system satisfies $\overline{P^{\mathrm{eff}}}=P_{\mathrm{ext}}$,
where $P_{\mathrm{ext}}$ is the peripheral pressure, assumed uniform.  An isolated monolayer under conditions of zero external loading must satisfy $P_{\mathrm{ext}}=0$, i.e.
\begin{equation}
    \label{eq:sero_stress_condition}
    \sum_{\alpha=1}^{N_c} A_\alpha P^{\mathrm{eff}}_\alpha = 0,
\end{equation}
allowing cells to be grouped into those that are under net tension ($P^{\text{eff}}_\alpha>0$) and net compression ($P^{\text{eff}}_\alpha<0$).
The deviatoric component of the tissue-level stress,
\begin{equation}
	\label{eq:tissue_isotropy}
	\bar{\boldsymbol{\sigma}}^{\mathcal{M}} = \frac{1}{A^{\mathcal{M}}} \sum_\alpha (T_\alpha + \mu \dot{L}_\alpha) L_\alpha \mathsf{J}_\alpha,
\end{equation}
has eigenvalues $\pm\xi$, where $\xi = \sqrt{\det{ (\boldsymbol{\sigma}^{\mathcal{M}}) } - \mathrm{Tr}( \boldsymbol{\sigma}^{\mathcal{M}} )^2 }$.  These quantify the tissue-level shear stress, and provide a measure of the spatial anisotropy of the monolayer \cite{kraynik2003}. 

The expression for tissue-level stress $\boldsymbol{\sigma}^{\mathcal{M}}=-\overline{P^{\mathrm{eff}}} \mathsf{I} + \bar{\boldsymbol{\sigma}}^{\mathcal{M}}$ using (\ref{eq:glob_pre}, \ref{eq:tissue_isotropy}) extends results derived in \citep{nestorbergmann2017, sugimura2013, xu2016oriented} to account for viscous resistance to area and perimeter changes, although it does not include additional dissipative stresses associated with neighbour exchanges or extrusion of very small cells.  An alternative derivation of $\boldsymbol{\sigma}^{\mathcal{M}}$ directly from $U$ and $\Phi$ is provided in Appendix~\ref{sec:mappings_derivation}, where we show that $\boldsymbol{\sigma}^{\mathcal{M}}:\dot{\mathsf{E}}=-(\dot{U}+\Phi)=0$ for a small-amplitude homogeneous strain $\mathsf{E}$; larger deformations, leading to neighbour exchanges, are therefore needed in order to change the system's internal energy.  The equilibria reported below are however unaffected by the choice of dissipation in that they are always local minima of $U(\mathbf{R}^1, \dots,\mathbf{R}^{N_v})$. 

\subsection{Simulation methodology}

Simulations were generated using the methodology outlined in \cite{nestorbergmann2017}.  Initial distributions of cell centres were generated using a Mat\'ern type II random sampling process.  Cell edges and vertices were formed by constructing a Voronoi tessellation about the seed points, imposing periodic boundary conditions in a square domain.  The system was then relaxed towards the nearest energy minimum.  Following the initialisation of the disordered geometry, a series of isotropic expansions or contractions were imposed until \eqref{eq:sero_stress_condition} was satisfied within a prescribed tolerance.  Stretching deformations were imposed by mapping all vertices and the domain boundary by an affine transformation and then allowing the system to relax.

The effective pressure (\ref{eq:peff}) of a regular hexagon at equilibrium is given by
\begin{equation}
    P^{\text{eff}}_6 = A - 1 + \frac{\Gamma\mu_6^{2}}{2} + \frac{\Lambda\mu_6}{4\sqrt{A}},
    \label{eq:p6}
\end{equation}
where $A$ is the area of the hexagon.  We define $A_6^*(\Gamma$, $\Lambda)$ to satisfy $P^{\text{eff}}_6(A_6^*)=0$, where the hexagon is stress free, identifying $\sqrt{A_6^*}$ as a length scale.  During relaxation, T1 transitions (neighbour exchanges) were performed on edges with length less than $0.1\sqrt{A_6^*}$.  3-sided cells with area less than $0.3A_6^*$ were removed via a T2 transition (extrusion).  \citet{fletcher2014vertex} and \citet{spencer2017} outline refined treatments of these deformations.

Our focus is primarily on mechanical properties of a monolayer across region II in Figure~\ref{fig:geometry}.  For comparison with experiments we adopt parameters fitted previously to our experimental system \cite{nestorbergmann2017}, namely $(\Lambda,\Gamma)=(-0.259, 0.172)$, acknowledging the imperfection of the fit and the inherent challenges of parameter estimation in this system \cite{kursawe2017}.

\subsection{Experimental methods}
\label{sec:expmeth}

Our \textit{Xenopus} embryonic animal cap preparation, stretch assay and imaging protocol are described in Appendix~\ref{sec:appexp}.  Briefly, explants from stage 10 \textit{Xenopus laevis} embryos were cultured on a fibronectin PDMS membrane.  The tissue layer is three cell layers thick; the basal cells attached to the membrane while the apical cells were imaged with confocal microscopy.  {\color{red}The apical cells were not in direct contact with the PDMS membrane, shielding them from influences such as substrate-mediated integrin activation or focal adhesion formation \citep{xu2014}.}  Uniaxial in-plane stretching of the rectangular PDMS membrane (with a strain imposed on two opposite lateral boundaries and no stress imposed on the other two) deformed the tissue layer, with strains transmitted from the membrane to the apical layer via the basal cells.  Using GFP-$\alpha$-tubulin cell cortex and cherry-histone nuclear markers, the images of the apical cells were manually segmented and cell boundaries were approximated as polygons using a Python script.  The effective tissue pressure $\overline{P^{\text{eff}}}$ of the unstretched monolayer was presumed to be zero, allowing the preferred area parameter $A_0^*$ to be calculated  following \cite{nestorbergmann2017_bio} using fitted parameters $(\Lambda,\Gamma)=(-0.259,0.172)$ from \cite{nestorbergmann2017}.  $A_0^*$ was assumed to be unchanged when the epithelium was stretched, allowing $P_\alpha^{\text{eff}}$ of individual cells to be estimated.  The principal axis of stress for individual cells was identified using the shape tensor based on each cell's vertex locations, as described in \cite{nestorbergmann2017}. 


\section{Perturbing tissue structure using deformations}

\subsection{Stretching embryonic epithelium}

Figure~2A shows apical cell boundaries, rendered as polygons, within a \textit{Xenopus} embryonic epithelium in the unstretched configuration.  Cell shapes are used to infer the relative isotropic stress $P_\alpha^{\mathrm{eff}}$ as described in Sec.~\ref{sec:expmeth}.  Cells estimated to be under net tension ({\color{red}darker}, with $P_\alpha^{\mathrm{eff}} \geq 0$) and under net compression ({\color{red}lighter}, with $P_\alpha^{\mathrm{eff}}<0$) appear in roughly equal proportions, with their orientations distributed approximately isotropically (Figure~2C).  The stress field is strongly heterogeneous at the single-cell level: cells under tension (compression) generate a primarily contractile (expansive) stress along their long (short) axis.  Figure~2B shows the same epithelium after an application of an instantaneous 35\% uniaxial stretch (in the horizontal direction) to the PDMS membrane; the finite thickness of the tissue means that the apical layer experiences a lower level of strain.   The isotropic distribution of apical cells in the undeformed state is disrupted by stretching: cells under net tension ({\color{red}darker}) tend to align their principal axis of stress with the direction of the global stretch, whereas cells under net compression ({\color{red}lighter}) tend to align their principal axis of stress vertically (Figure 2D).  Stretch therefore induces anisotropy and a degree of order to the monolayer, consistent with previous observations {\color{red}\citep{sugimura2013, nestorbergmann2017_bio, wyatt2015, harris2012, xu2015}}.  

\begin{figure} 
\includegraphics[width=0.85\textwidth]{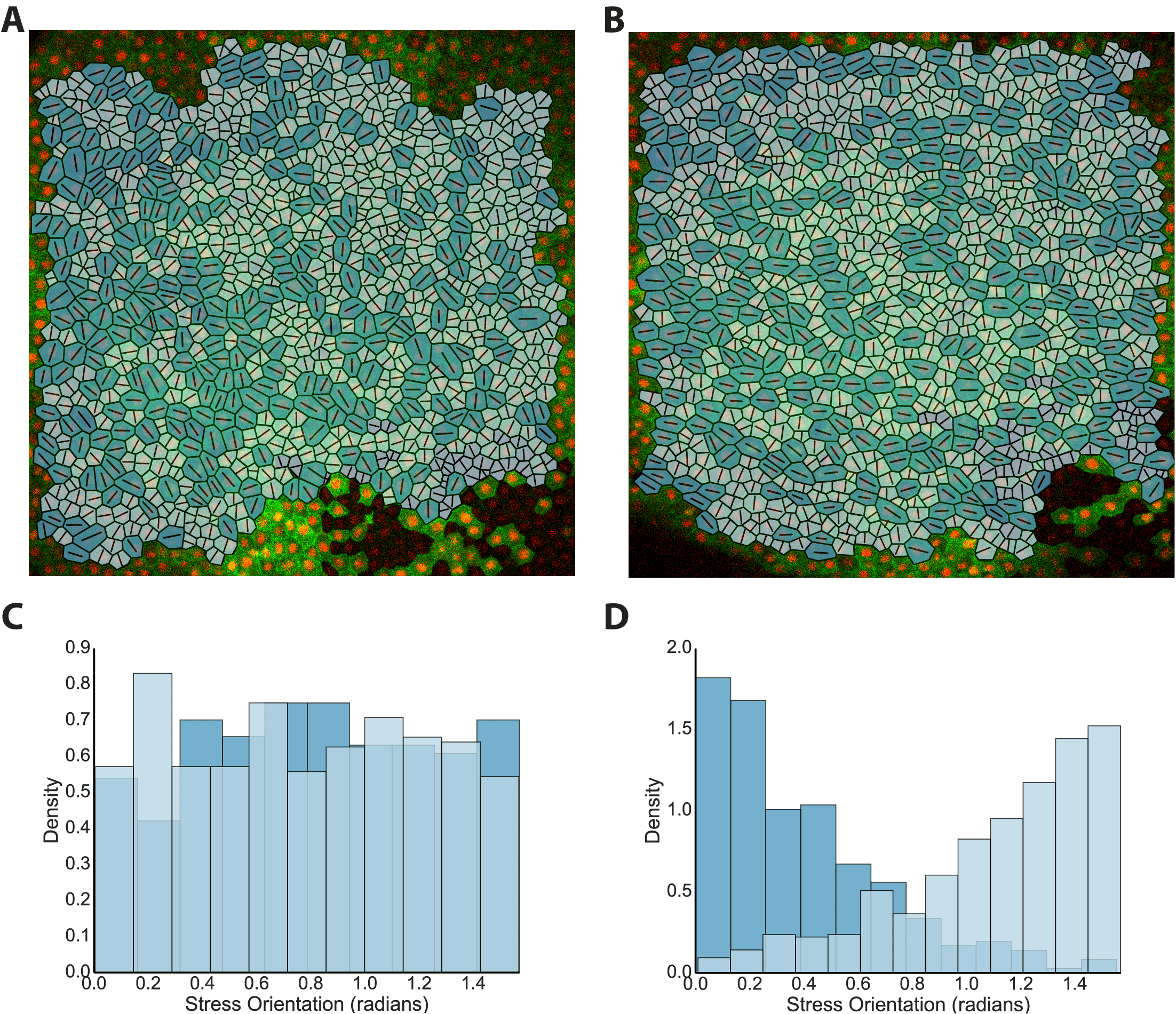}
\label{fig:expt}
\caption{ \textbf{A}.  Epithelial apical layer of a \textit{Xenopus laevis} animal cap, showing 801 cells rendered as polygons superimposed on the original microscopy image.  $P_\alpha^{\text{eff}}$ for each cell was calculated assuming $\overline{P^{\text{eff}}}=0$ and $(\Lambda,\Gamma)= (-0.259,0.172)$. Line segments indicate the principal axis of shape and stress for each cell. {\color{red} Darker (lighter)} cells have $P_\alpha^{\text{eff}} > 0~(< 0)$ and exert a net inward (outward) force along each line segment. \textbf{B.} The apical layer in \textbf{A} following a 35\% instantaneous uniaxial stretch (horizontal) of the membrane beneath the basal cells, resulting in an $19.67\pm1.91\%$ (95\% confidence interval) uniaxial stretch of the apical cells. \textbf{C, D.}  Histograms showing {\color{red}frequency density of} orientation of the principal axis of stress for cells under tension ({\color{red}darker}) and compression ({\color{red}lighter}), for apical layers given in \textbf{A} (corresponding to \textbf{C}) and \textbf{B} (corresponding to \textbf{D}); {\color{red}bin areas integrate to unity}. 
Bin size was selected using the Freedman--Diaconis rule.}
\end{figure}

\subsection{Simulated tissue stretching}

\begin{figure*} 
	\centering
	\includegraphics[width=0.85\textwidth]{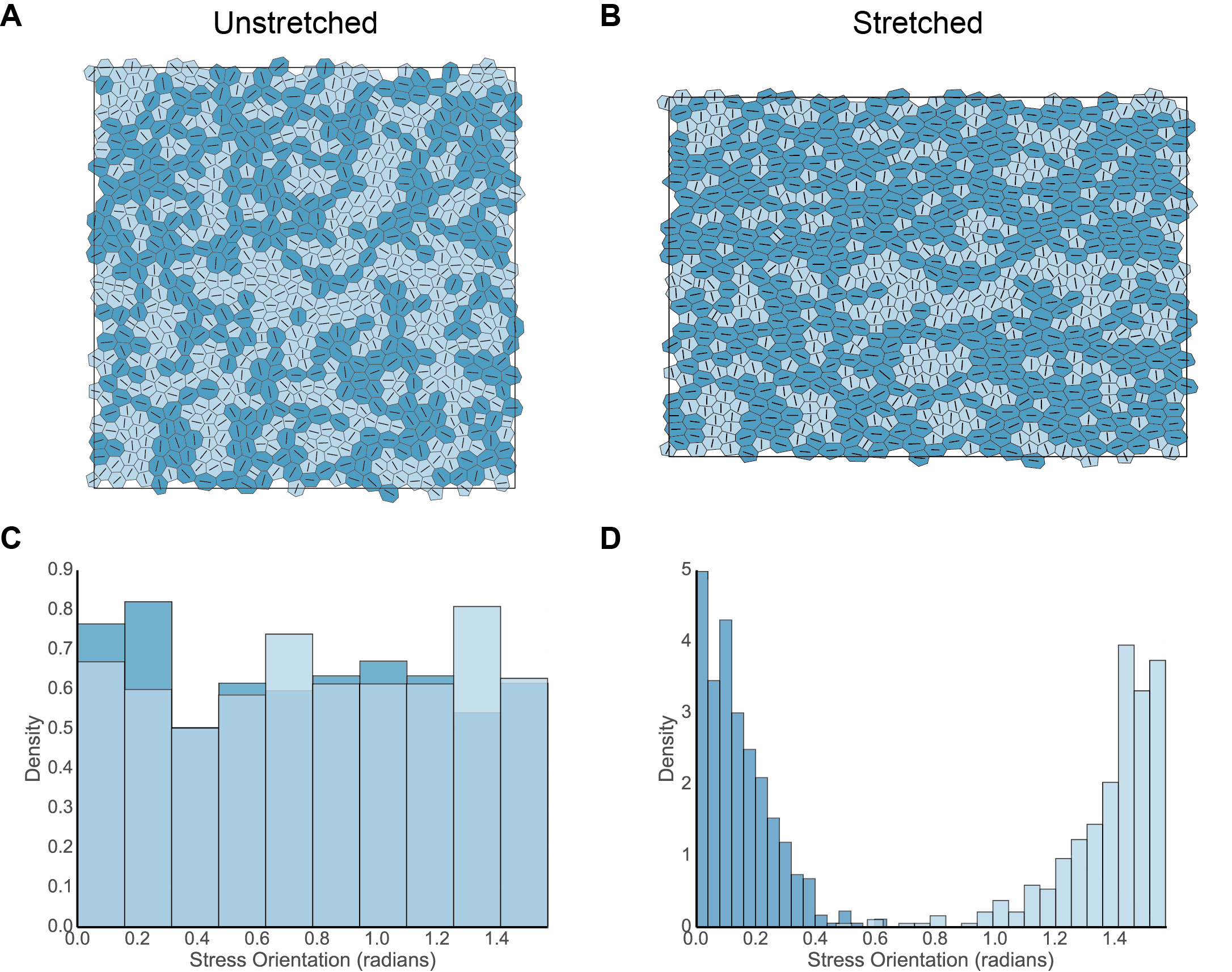}
	\caption{\textbf{A}. A simulation of a representative monolayer realisation satisfying $\overline{P^{\mathrm{eff}}}=0$ with 800 cells, for $(\Lambda,\Gamma) = (-0.259, 0.172)$ (see Figure~\ref{fig:geometry} for location in parameter space).  Cell shadings and line segments follow the scheme in Figure~2.  \textbf{B}.  The monolayer in \textbf{A} following a 20\% area-preserving uniaxial stretch and subsequent relaxation.  \textbf{C\&D.} Histograms showing orientation of the principal axis of stress for cells under tension ({\color{red}darker}) and compression ({\color{red}lighter}), for monolayers given in \textbf{A} (corresponding to \textbf{C}) and \textbf{B} (corresponding to \textbf{D}).  Bin size was selected using the Freedman--Diaconis rule.}
    \label{fig:example_stretch}
\end{figure*}

We mimic these observations using simulations, seeking to characterize the mechanical properties of the deformed monolayer.  {\color{red} We describe the  passive response to stretch, ignoring additional mechanosensitive effects of the kind described at the single-cell level by Xu \hbox{et al.} \cite{xu2016}.}  Figure~\ref{fig:example_stretch} shows the result of performing a $20\%$ area-preserving stretch on a simulated monolayer, with the mapping of vertices 
\begin{equation}
\mathbf{R}^j = (R^j_x,R^j_y) \rightarrow ((1+\lambda) R^j_x, R^j_y/(1+\lambda))
\label{eq:mapping}
\end{equation}
where $\lambda=0.2$, followed by relaxation to equilibrium via T1 transitions.  In the undeformed state, the cell orientations are approximately isotropic (Figure~\ref{fig:example_stretch}C), and roughly equal proportions of cells are under net tension and net compression (Figure~\ref{fig:example_stretch}A).   Stretching increases the proportion of cells under net tension (Figure~\ref{fig:example_stretch}B), inducing a striking degree of orientational order.  As in Figure~2D, cells under net tension tend to align their principal axis of stress with the direction of stretch, whereas cells under net compression tend to align their principal axis of stress perpendicularly (Figure~\ref{fig:example_stretch}D).  The ordering is more extreme than in experiments; it is likely that for this large deformation, refinements of the functional form of the mechanical energy $U$ (leading to linear pressure and tension relations in (\ref{eq:pt})) are needed to make the model quantitatively accurate.

\begin{figure} 
	\centering
	\includegraphics[width=0.45\textwidth]{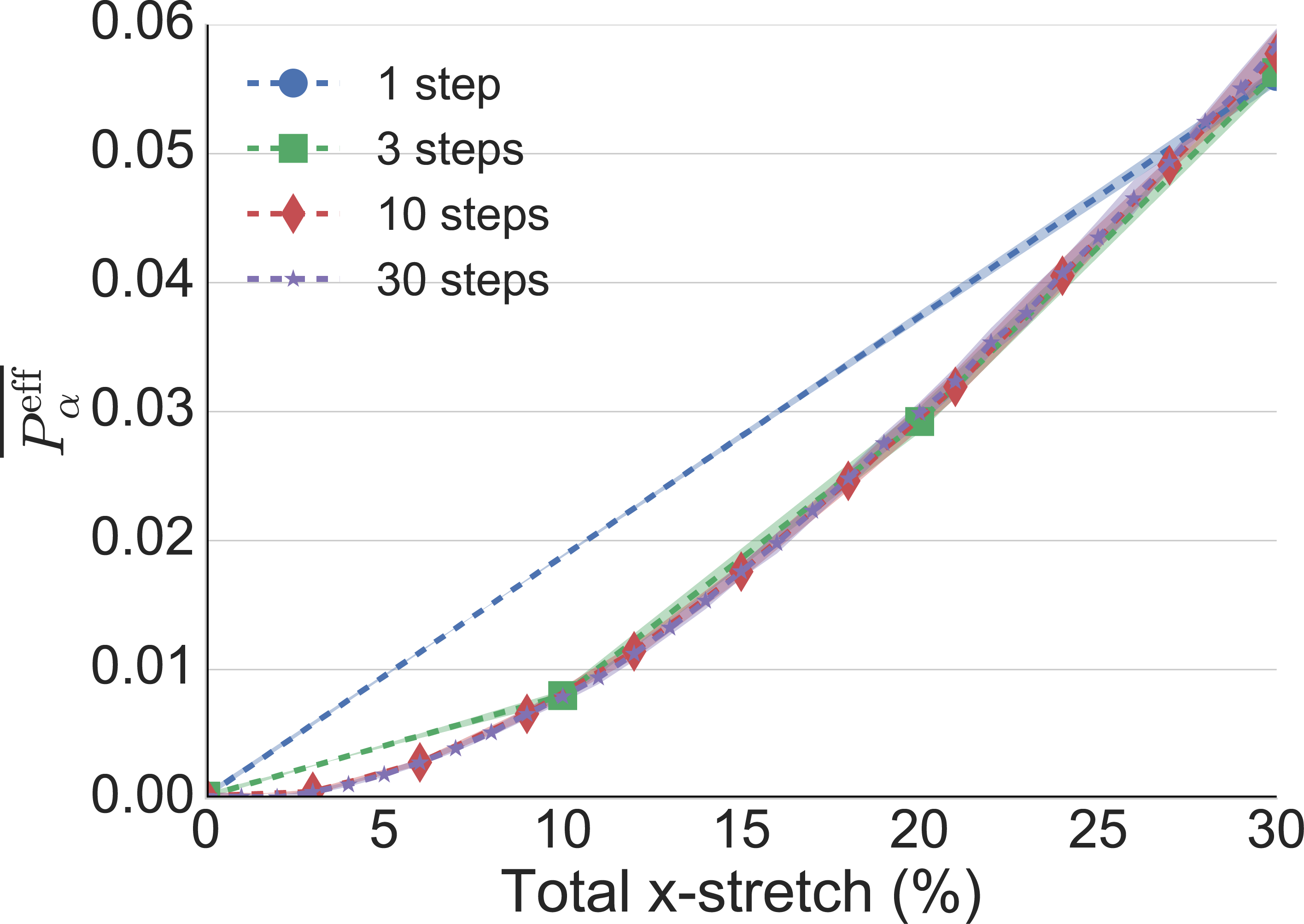}
	\includegraphics[width=0.45\textwidth]{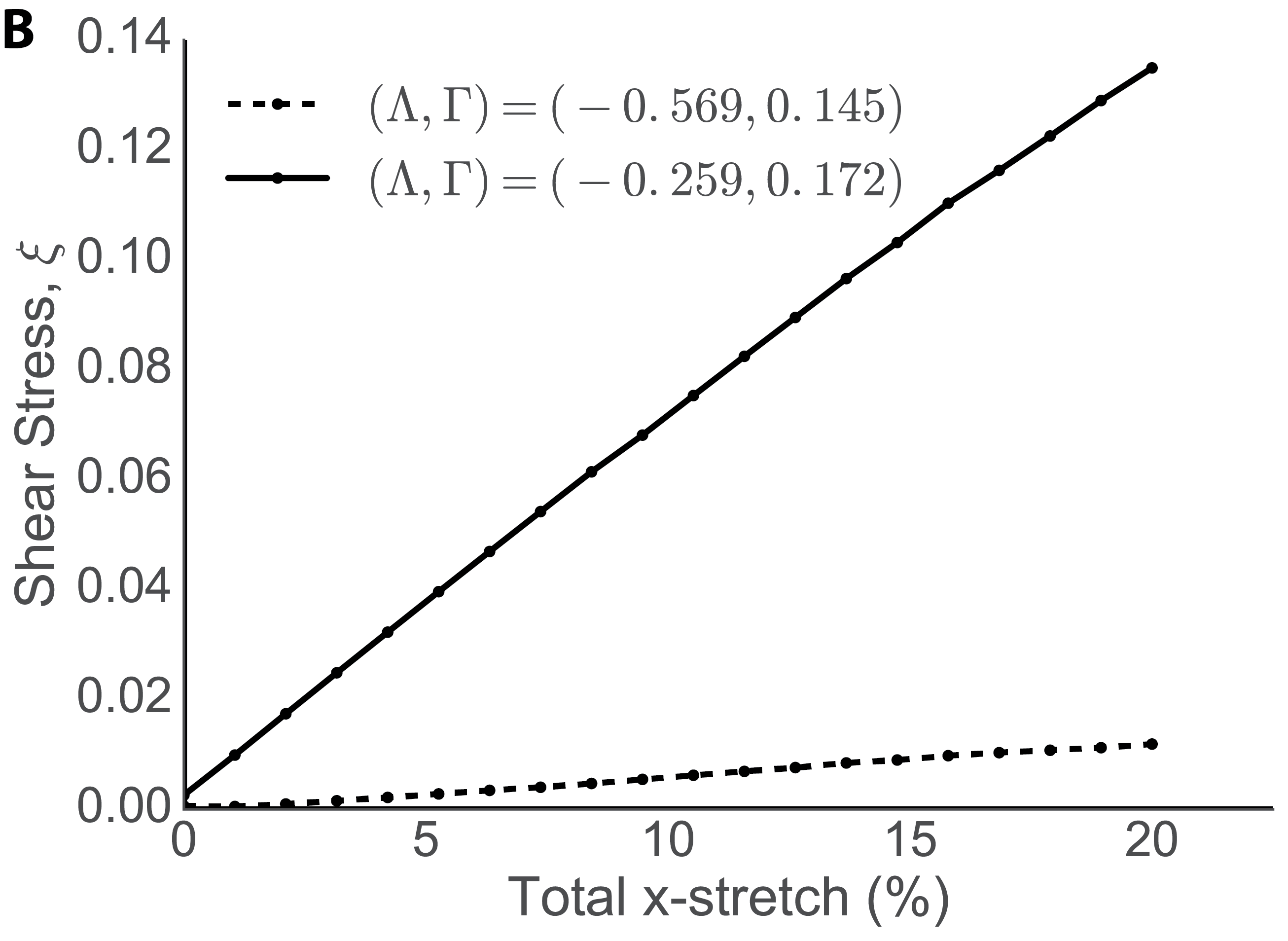}
	\caption{\textbf{A}. The effect of incremental stretch on effective tissue pressure.  The tissue shown in Figure~\ref{fig:example_stretch}A was subjected to a 30\% area-preserving uniaxial stretch in a varying number of steps (straight lines are drawn between datapoints).  The total stretch was divided into equally spaced increments and the tissue was relaxed after every stretch.  The tissue starts at $\overline{P^{\mathrm{eff}}}=0$ and ends at approximately $\overline{P^{\mathrm{eff}}}=0.57$, regardless of how many steps were used.  Translucent shading indicates 95\% confidence intervals over the 5 simulations. 
    \label{fig:stress_vs_stretch}
\textbf{B}.  Shear stress, $\xi$, versus magnitude of stretch with $(\Lambda,\Gamma) = (-0.259, 0.172)$ (solid; the monolayer given in Figure~\ref{fig:example_stretch}) and $(\Lambda,\Gamma) = (-0.569, 0.145)$ (dashed).  See Figure~\ref{fig:geometry}B for locations in parameter space.  Each datapoint represents an instantaneous stretch performed on the same initial isotropic monolayer satisfying $P_{\mathrm{ext}}=0$.  The monolayers were relaxed to equilibrium following stretch. }
\end{figure}

In addition, the effective tissue pressure $\overline{P^{\mathrm{eff}}}$, measuring the isotropic component of stress in the monolayer (\ref{eq:peff}), increases with the degree of stretch (Figure~\ref{fig:stress_vs_stretch}A).  $\overline{P^{\mathrm{eff}}}$ is not significantly affected by the number of steps in which stretching is done, even when the monolayer is relaxed after every step  (Figure~\ref{fig:stress_vs_stretch}A).  In the linear regime ($\lambda \ll 1$), (\ref{eq:mapping}) is a pure shear deformation, leading to negligible increase in $\overline{P^{\mathrm{eff}}}$ for small stretches;  however a nonlinear response emerges at larger amplitudes.  Figure~\ref{fig:stress_vs_stretch}B demonstrates that the tissue shear stress, $\xi$ (see (\ref{eq:tissue_isotropy})), induced by individual stretches of increasingly large amplitude from an unstressed isotropic inital condition (with subsequent relaxation via T1 transitions), increases with stretch magnitude.  This holds for both the monolayer given in Figure~\ref{fig:example_stretch} and an example closer to the region I/II boundary (Figure~\ref{fig:geometry}B) where $(\Lambda,\Gamma) = (-0.569, 0.145)$.  The latter has a significantly smaller slope, indicating much lower resistance to shear in this region of parameter space.   {\color{red}For reference, the anisotropy in these examples measured by a more traditional order parameter is illustrated in Appendix~\ref{app:c}.  The present model does not account for stress relaxation that may arise under large strains via cell division, as illustrated in simulations by \cite{xu2015}.}

Given that stretching induces a strong degree of anisotropy in the tissue, we now assess how this ordering influences further tissue deformation.  An initially isotropic tissue is uniaxially stretched via (\ref{eq:mapping}) and relaxed to equilibrium, as above.  This deformed (\hbox{i.e.} pre-stretched) configuration has some pre-stress $\boldsymbol{\sigma}^{0}$.  A further small-amplitude homogeneous strain, $\mathsf{I} \rightarrow \mathsf{I} + \mathsf{E}$, changes the global stress as  $\boldsymbol{\sigma}^{0} \rightarrow \boldsymbol{\sigma}^{0} + \Delta \boldsymbol{\sigma}^{\mathcal{M}} $.  We evaluate the perturbation stresses  $\Delta \boldsymbol{\sigma}^{\mathcal{M}}$ arising from unidirectional strains $\mathsf{E}^X= \mathrm{diag}(\lambda,0)$ and $\mathsf{E}^Y=\mathrm{diag}(0,\lambda)$ directly from (\ref{eq:global_prestress}), noting that each deformation combines expansion and shear, \hbox{e.g.} 
\begin{equation}
    \label{eq:composite_strain}
    \mathsf{E}^X = \begin{bmatrix} \lambda & 0 \\ 0 & 0  \end{bmatrix} = \begin{bmatrix} \tfrac{\lambda}{2} & 0 \\ 0 & \tfrac{\lambda}{2}  \end{bmatrix} + \begin{bmatrix} \tfrac{\lambda}{2} & 0 \\ 0 & -\tfrac{\lambda}{2} \end{bmatrix}.
\end{equation}
It follows that both the shear and bulk elasticity of the tissue will contribute to the induced perturbation stress.

\begin{figure} 
	\centering
	\includegraphics[width=0.55\textwidth]{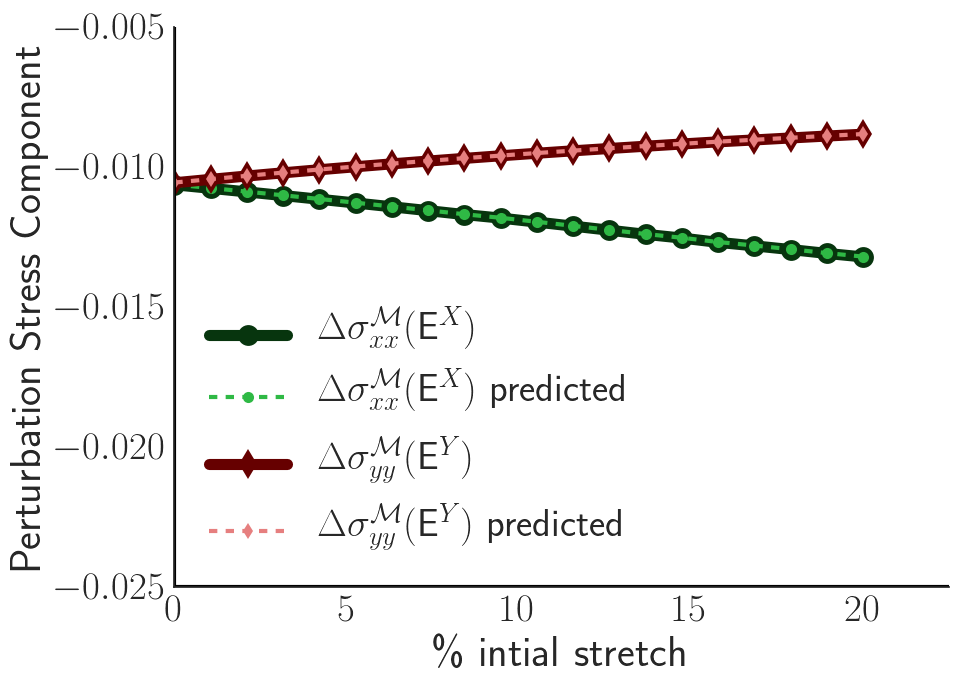}
	\caption{Perturbation stress response to small amplitude deformations in a pre-stretched monolayer.  The stretched monolayers used were the same as those for the solid line in Figure~\ref{fig:stress_vs_stretch}B, with $(\Lambda,\Gamma) = (-0.259, 0.172)$ and the magnitude of pre-stretch is indicated on the $x$-axis.  The equilibrium pre-stretched monolayers were subjected to small deformations in the $x$- ($\mathsf{E}^X = \mathrm{diag}(\lambda,0)$) and $y$- ($\mathsf{E}^Y = \mathrm{diag}(0,\lambda)$) directions, with $\lambda=0.01$.  The component of the perturbation stress tensor in the direction of stretch is indicated on the $y$-axis, with $\Delta \sigma^{\mathcal{M}}_{xx}(\mathsf{E}^X)$ giving the $x$-directed stress following $\mathsf{E}^X$ (lower line) and $\Delta \sigma^{\mathcal{M}}_{yy}(\mathsf{E}^Y)$ giving the $y$-directed stress following $\mathsf{E}^Y$ (upper line). Solid lines indicate values directly evaluated using \eqref{eq:global_prestress} and dashed lines are predicted values using \eqref{eq:tissue_pert_stress}. }
    \label{fig:pert_deformations}
\end{figure}

Figure~\ref{fig:pert_deformations} plots two components of the perturbation stress against the magnitude of pre-stretch, for the tissue shown in Figure~\ref{fig:example_stretch}A, when subjected to additional 1\% strains in the $x$- and $y$- directions.  When the tissue is subject to weak initial stretch, subsequent perturbation stresses are almost equal ($\Delta \sigma^{\mathcal{M}}_{xx}(\mathsf{E}^X) \approx \Delta \sigma_{yy}^{\mathcal{M}}(\mathsf{E}^Y)$), indicating that the tissue is mechanically isotropic.  However, under increased pre-strain, we see that $\Delta \sigma^{\mathcal{M}}_{xx}(\mathsf{E}^X) > \Delta \sigma^{\mathcal{M}}_{yy}(\mathsf{E}^Y)$, indicating increased anisotropy.  The figure demonstrates how the tissue can be preferentially stiffened in one direction by imposition of pre-stretch.   Correspondingly, of the membrane elements shown in Figure~\ref{fig:example_stretch}B, a greater net membrane length is oriented in the $x$ direction (resisting further stretch) in comparison to the $y$ direction. 

\section{Deriving the stiffness tensor for anisotropic tissues}

We have demonstrated that anisotropic deformations induce ordering in a tissue, leading to an anisotropic response to external loading reminiscent of an orthotropic material.  We now evaluate directly the stiffness and damping tensors $\mathsf{C}(\boldsymbol{\sigma}^0)$ and $\mathsf{D}(\boldsymbol{\sigma}^0)$ appearing in the linearized relation connecting perturbation strain and  perturbation stress $\Delta\boldsymbol{\sigma} =  -(\mathsf{C}: \mathsf{E}+\mathsf{D}:\dot{\mathsf{E}})$ for some imposed strain $\mathsf{E}$.  This small-amplitude relation does not account directly for stress relaxation via neighbour exchanges, although these can influence the pre-stress $\boldsymbol{\sigma}_0$ of the base state.

\subsection{The perturbation stress of a single cell}

We begin by considering a single cell with stress given by (\ref{eq:stres1}).  From a stationary state with pre-stress 
\begin{equation}
	\label{eq:prestress}
	\boldsymbol{\sigma}^o_\alpha = -(A_\alpha-1)\mathsf{I} - \frac{T_\alpha L_\alpha}{A_\alpha} \mathsf{Q_\alpha},
\end{equation}
we impose a small-amplitude, homogenous, symmetric, time-varying strain $\mathsf{E}$, such that position vectors transform as $\mathbf{R} \rightarrow \mathbf{R} + \mathsf{E}\cdot\mathbf{R}$.  Linearizing in the small strain amplitude yields the mappings (Appendix~\ref{sec:mappings_derivation})
\begin{equation}
	\begin{gathered}
	\label{eq:mapings}
     L_\alpha \rightarrow L_\alpha(1 + \mathsf{Q}_\alpha : \mathsf{E} ), \quad
	A_\alpha \rightarrow A_\alpha(1 + \mathrm{Tr}(\mathsf{E})), \\
    T_\alpha \rightarrow T_\alpha + \Gamma L_\alpha \mathsf{Q}_\alpha : \mathsf{E}, \quad 
	L_\alpha\mathsf{Q}_\alpha \rightarrow L_\alpha\mathsf{Q}_\alpha + L_\alpha\mathsf{B}_\alpha : \mathsf{E}, 
	\end{gathered}
\end{equation}
where $\mathsf{B}_\alpha$ is a fourth-order tensor (\ref{eq:B}) satisfying
\begin{equation}
    \mathsf{B}_\alpha:\mathsf{E} = \frac{1}{L_\alpha} \sum_{i=0}^{Z_\alpha-1} l^i_\alpha \left[ \hat{\mathbf{t}}_{\alpha}^{i} \otimes (\mathsf{E}\cdot \hat{\mathbf{t}}_{\alpha}^{i}) + (\mathsf{E}\cdot \hat{\mathbf{t}}_{\alpha}^{i} )\otimes \hat{\mathbf{t}}_{\alpha}^{i} - \hat{\mathbf{t}}_{\alpha}^{i} \otimes \hat{\mathbf{t}}_{\alpha}^{i} (\hat{\mathbf{t}}_{\alpha}^{i} \cdot\mathsf{E}\cdot \hat{\mathbf{t}}_{\alpha}^{i}) \right]
    \label{eq:BE}
\end{equation}
so that $\mathrm{Tr}(\mathsf{B}_\alpha : \mathsf{E}) = \mathsf{Q}_\alpha : \mathsf{E}$.  Dynamic area and perimeter changes are coupled to $\mathsf{E}(t)$ via $\dot{A}_\alpha=A_\alpha\mathrm{Tr}(\dot{\mathsf{E}})$ and $\dot{L_\alpha}=L_\alpha\mathsf{Q}_\alpha:\dot{\mathsf{E}}$.  

Writing the stress following the deformation as $\boldsymbol{\sigma}_\alpha = \boldsymbol{\sigma}_\alpha^o + \Delta\boldsymbol{\sigma}_\alpha$, we use \eqref{eq:mapings} and \eqref{eq:stres1} to give, for a monolayer at equilibrium,
\begin{equation}
	\begin{split}
	\boldsymbol{\sigma}_\alpha &= -(A_\alpha-1)\mathsf{I} - A_\alpha\;\mathrm{Tr}(\mathsf{E}) \mathsf{I} - \frac{T_\alpha + \Gamma L_\alpha \mathsf{Q}_\alpha : \mathsf{E}}{A_\alpha(1 + \mathrm{Tr}(\mathsf{E}))} (L_\alpha\mathsf{Q}_\alpha + L_\alpha \mathsf{B}_\alpha : \mathsf{E}) \\
	&\approx -(A_\alpha-1)\mathsf{I} - A_\alpha\;\mathrm{Tr}(\mathsf{E}) \mathsf{I} 
      - \frac{1}{A_\alpha}(T_\alpha + \Gamma L_\alpha \mathsf{Q}_\alpha : \mathsf{E})(1 - \mathrm{Tr}(\mathsf{E})) (L_\alpha\mathsf{Q} + L_\alpha \mathsf{B}_\alpha : \mathsf{E}) + \dots \\
	&\approx -(A_\alpha-1)\mathsf{I} - \frac{T_\alpha L_\alpha}{A_\alpha} \mathsf{Q}_\alpha - A_\alpha\;\mathrm{Tr}(\mathsf{E})\mathsf{I} - \frac{\Gamma L_\alpha^2}{A_\alpha} (\mathsf{Q}_\alpha : \mathsf{E}) \mathsf{Q}_\alpha 
        + \frac{T_\alpha L_\alpha}{A_\alpha} \left[ \mathrm{Tr}(\mathsf{E}) \mathsf{Q}_\alpha - \mathsf{B}_\alpha : \mathsf{E} \right] + \dots
	\end{split}
\end{equation}
to first order.  Including time-dependent terms, the perturbation stress is therefore
\begin{equation}
	\Delta\boldsymbol{\sigma}_\alpha = - A_\alpha\;\left[\mathrm{Tr}(\mathsf{E})+\gamma\mathrm{Tr}(\dot{\mathsf{E}})\right]\mathsf{I} - \frac{L_\alpha^2}{A_\alpha} \mathsf{Q}_\alpha :(\Gamma  \mathsf{E}+\mu \dot{\mathsf{E}}) \mathsf{Q}_\alpha + \frac{T_\alpha L_\alpha}{A_\alpha} \left[ \mathrm{Tr}(\mathsf{E}) \mathsf{Q}_\alpha - \mathsf{B}_\alpha : \mathsf{E} \right].
\end{equation}
We can separate the isotropic and deviatoric contributions as
\begin{equation}
    \label{eq:pert_stress}
	\Delta\boldsymbol{\sigma}_\alpha = - \Delta{P}_\alpha^{\text{eff}} \mathsf{I} + \frac{T_\alpha L_\alpha}{A_\alpha} \Delta{\mathsf{J}_\alpha},
\end{equation}
where $\Delta{P}_\alpha^{\text{eff}}$ is the perturbation effective pressure
\begin{equation}
	\Delta{P}_\alpha^{\text{eff}} = \left(A_\alpha-\frac{T_\alpha L_\alpha}{2A_\alpha}\right) \mathrm{Tr}(\mathsf{E}) +\gamma A_\alpha \mathrm{Tr}(\dot{\mathsf{E}})+ \frac{\Gamma L_\alpha^2}{A_\alpha}\left( 1 - \frac{L_0}{2L_\alpha} \right) \mathsf{Q}_\alpha : \mathsf{E} + \frac{\mu L_\alpha^2 }{2A_\alpha} \mathsf{Q}_\alpha:\dot{\mathsf{E}},
	\label{eq:dpef}
\end{equation}
and the traceless contribution characterizing perturbation shear is 
\begin{equation}
    \begin{split}
T_\alpha \Delta{\mathsf{J}}_\alpha &=  \left[-\frac{T_\alpha}{2} \mathrm{Tr}(\mathsf{E}) + {\Gamma L_\alpha}\left( 1 - \frac{L_0}{2L_\alpha} \right) \mathsf{Q}_\alpha : \mathsf{E} +\frac{\mu L_\alpha}{2} \mathsf{Q}_\alpha:\dot{\mathsf{E}}\right] \mathsf{I} \\
      & \qquad \qquad \qquad \qquad
     + \left[ T_\alpha \mathrm{Tr}(\mathsf{E}) - { L_\alpha \mathsf{Q}_\alpha : (\Gamma \mathsf{E}+\mu\dot{\mathsf{E}})} \right] \mathsf{Q} _\alpha - T_\alpha \mathsf{B}_\alpha : \mathsf{E}.
     \end{split}
     \label{eq:dtj}
\end{equation}

\subsection{The perturbation stress of the tissue}

Applying $\mathsf{E}$ to the entire monolayer, the global stress transforms as $\boldsymbol{\sigma}^{\mathcal{M}} \rightarrow \boldsymbol{\sigma}^{\mathcal{M}} + \Delta \boldsymbol{\sigma}^{\mathcal{M}}$ so that \eqref{eq:global_prestress} becomes
\begin{equation}
\label{eq:27}
	\begin{split}
	A^{\mathcal{M}}&(1+\mathrm{Tr}(\mathsf{E}))( \boldsymbol{\sigma}^{\mathcal{M}} + \Delta \boldsymbol{\sigma}^{\mathcal{M}} ) 
	= \sum_\alpha^{N_c} A_\alpha(1+\mathrm{Tr}(\mathsf{E})) \left( \boldsymbol{\sigma}_\alpha^o + \Delta \boldsymbol{\sigma}_\alpha \right) \\
	& =  \sum_\alpha^{N_c} A_\alpha \left[ -P^{\text{eff}}_\alpha (1+\mathrm{Tr}(\mathsf{E})) - \Delta{P}^{\text{eff}}_\alpha \right] \mathsf{I} 
        + T_\alpha L_\alpha \left[ \mathsf{J}_\alpha(1+\mathrm{Tr}(\mathsf{E})) + \Delta{\mathsf{J}}_\alpha \right] +\dots,
	\end{split}
\end{equation}
neglecting terms quadratic in $\mathsf{E}$.  Linearizing the left-hand side of (\ref{eq:27}), 
the global perturbation stress is given by
\begin{equation}
	\begin{split}
	\label{eq:tissue_pert_stress}
	\Delta \boldsymbol{\sigma}^{\mathcal{M}} &= \frac{1}{A^{\mathcal{M}}} \sum_\alpha^{N_c} A_\alpha \Delta \boldsymbol{\sigma}_\alpha 
	= \frac{1}{A^{\mathcal{M}}} \sum_\alpha^{N_c} \left[ -A_\alpha \Delta{P}^{\text{eff}}_\alpha \mathsf{I} + T_\alpha L_\alpha \Delta{\mathsf{J}}_\alpha \right].
	\end{split}
\end{equation}
Thus the effective perturbation tissue pressure is
\begin{equation}
    \overline{\Delta{P}^{\mathrm{eff}}} = \frac{1}{A^{\mathcal{M}}} \sum_\alpha^{N_c} A_\alpha \Delta{P}^{\text{eff}}_\alpha.
    \label{eq:opef}
\end{equation}
The predictions arising from \eqref{eq:tissue_pert_stress} in pre-stretched monolayers being subjected to small-amplitude strains are tested in Figure~\ref{fig:pert_deformations}, showing good agreement with direct stress computations.  Thus, for a given $\mathsf{E}$, the stiffness matrix, $\mathsf{C}$, can be evaluated directly from the terms proportional to $\mathsf{E}$ in (\ref{eq:tissue_pert_stress}) and its viscous analogue $\mathsf{D}$ from terms proportional to $\dot{\mathsf{E}}$:
{\color{red}
\begin{subequations}
\label{eq:cd}
\begin{align}
\mathsf{C}&=\frac{1}{A^\mathcal{M}} \sum_\alpha^{N_c}\left[  A_\alpha^2 \mathsf{I}\otimes \mathsf{I} +\Gamma L_\alpha^2 \mathsf{Q}_\alpha\otimes \mathsf{Q}_\alpha +L_\alpha T_\alpha (\mathsf{B}_\alpha-\mathsf{Q}_\alpha\otimes \mathsf{I})\right], \\
\mathsf{D}&=\frac{1}{A^\mathcal{M}} \sum_\alpha^{N_c}\left[ \gamma A_\alpha^2 \mathsf{I}\otimes \mathsf{I}+\mu L_\alpha^2 \mathsf{Q}_\alpha\otimes \mathsf{Q}_\alpha\right],
\end{align}
\end{subequations}
using the notation $\{\mathsf{A}\otimes\mathsf{B}\}_{ijkl}=A_{ij}B_{kl}$.
}
We now illustrate this in the special case of an initially unstressed disordered monolayer.

\subsection{Elastic moduli for a disordered isotropic monolayer}

\begin{figure*} 
	\centering
	\includegraphics[width=0.9\textwidth]{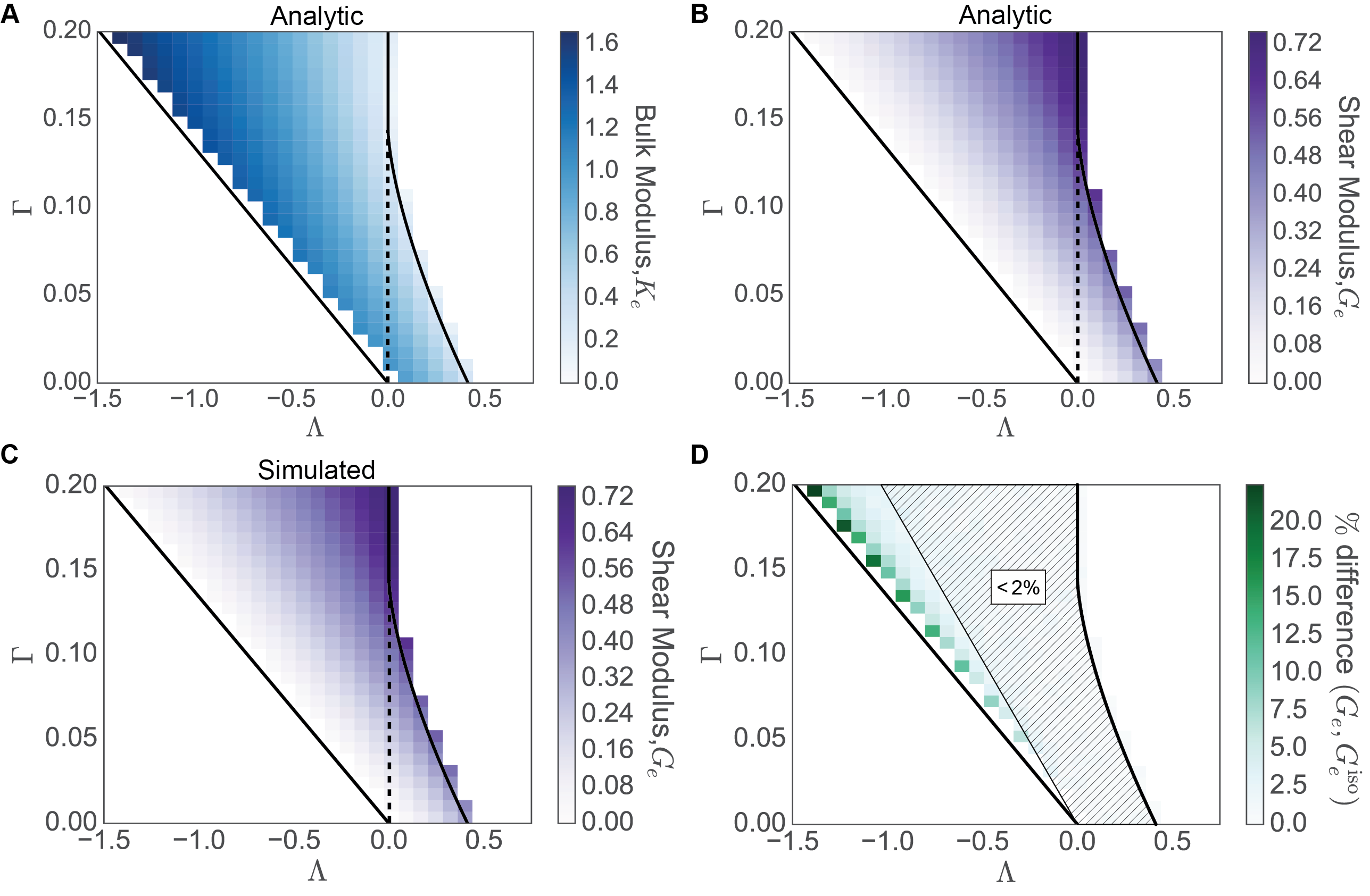}
	\caption{\textbf{A.} Heat map across discrete intervals of $(\Lambda,\Gamma)$ parameter space showing the value of the analytic bulk elastic modulus $K_e$ (calculated using \eqref{eq:bulk}) of a disordered isotropic monolayer with 800 cells.  \textbf{B.}  Equivalent plot of the analytic shear modulus (calculated using \eqref{eq:disordered_shearmod}).  \textbf{C.}  Computationally estimated shear modulus across parameter space.  The elastic shear modulus, $G_e$, was estimated from the global perturbation stress as $\Delta \sigma^{\mathcal{M}}_{xy}=-\kappa G_e$, following a 1\% simple-shear deformation ($\kappa=0.01$) on the simulated monolayers used in \textbf{B}.  Each datapoint is taken as an average from 5 realisations of a monolayer with 800 cells, with $P_\text{ext}=0$.  \textbf{D.} Percentage difference ($|G_e-G_e^{\mathrm{iso}}| / (0.5(G_e + G_e^{\mathrm{iso}}))$) between the exact modulus $G_e$ \eqref{eq:disordered_shearmod} and the approximation $G_e^{\mathrm{iso}}$ \eqref{eq:disordered_shearmod_iso}, across parameter space; the difference is below 2\% in the hatched region.   }
    \label{fig:disordered_bulk_shear_mod}
\end{figure*}

When the base state is a disordered isotropic monolayer at zero stress (satisfying (\ref{eq:sero_stress_condition}) with $\bar{\boldsymbol{\sigma}}^\mathcal{M}=\mathsf{0}$), we can derive bulk moduli by imposing  a small isotropic expansion with $\mathsf{E} = \lambda \mathsf{I}$, where $\lambda \ll 1$.  The deformation satisfies
\begin{equation}
	\label{eq:iso_stretch_idents}
	\mathrm{Tr}(\mathsf{E}) = 2\lambda, \qquad \mathsf{Q}_\alpha : \mathsf{E} = \lambda, \qquad \mathsf{B}_\alpha : \mathsf{E} = \lambda \mathsf{Q}_\alpha.
\end{equation}
Under an isotropic load, the deviatoric components of the perturbation stress vanish ($\sum_{\alpha=1}^{N_c} T_\alpha L_\alpha \Delta{\mathsf{J}}_\alpha = 0$).  Using \eqref{eq:iso_stretch_idents}, the bulk perturbation effective pressure (\ref{eq:opef}) is
\begin{equation}
\overline{\Delta P^{\mathrm{eff}}}=\frac{1}{A^{\mathcal{M}}} \sum_\alpha \left\{ \left( 2A_\alpha^2 + \frac{\Gamma L_0 L_\alpha}{2}\right) \lambda + \left( 2\gamma A_\alpha^2 + \frac{\mu L_\alpha^2}{2}  \right) \dot{\lambda} \right\}.
\label{eq:peffA}
\end{equation}
The bulk and cortical viscosities (appearing in the coefficient of $\dot{\lambda}$ in (\ref{eq:peffA})) contribute in a similar manner to $\overline{\Delta P^{\mathrm{eff}}}$ as the bulk and cortical stiffnesses, except via a nonlinear dependence on $L_\alpha$.  Recalling that $\Delta A^\mathcal{M} = A^\mathcal{M} \mathrm{Tr}(\mathsf{E})$, the bulk elastic modulus, $K_e$, can be derived from (\ref{eq:peffA}) with $\dot{\lambda}=0$ as
\begin{equation}
	\label{eq:bulk}
	K_e = A^\mathcal{M} \frac{ \overline{\Delta{P}^{\mathrm{eff}}}}{\Delta A^\mathcal{M}} 
    = \sum_{\alpha=1}^{N_c} \frac{1}{2A^\mathcal{M}} \left[ 2A_{\alpha}^2+\frac{\Gamma L_0 L_\alpha}{2} \right],
\end{equation}
in agreement with \cite{nestorbergmann2017}.  Figure~\ref{fig:disordered_bulk_shear_mod}A demonstrates how $K_e$ varies across parameter space, with the tissue becoming less resistant to isotropic deformation towards the region II/III boundary.  The tissue stress arises through area-weighting of cellular stress \eqref{eq:global_prestress}, leading to a nonlinear area dependence of bulk modulus on cell area in (\ref{eq:bulk}); thus when cells are substantially smaller than their target area (near the II/III boundary) the bulk modulus falls accordingly, approaching near-zero values.

For the shear moduli, we impose a small simple-shear deformation with $\mathsf{E} = \kappa \mathbf{e}_x \mathbf{e}_y$, where $\vert \kappa\vert \ll 1$ and $\mathbf{e}_x=(1,0), \mathbf{e}_y=(0,1)$ are the Cartesian coordinate basis, and seek $\Delta\boldsymbol{\sigma}_{xy}$.  This simple deformation satisfies $\mathrm{Tr}(\mathsf{E}) = 0$.  To evaluate $\mathsf{Q}_\alpha:\mathsf{E}$ and $\mathsf{B}_\alpha:\mathsf{E}$, we write $\hat{\mathbf{t}}_\alpha^{i} = \cos\theta_\alpha^i \mathbf{e}_x + \sin\theta_\alpha^i \mathbf{e}_y$, where $\theta_\alpha^i$ satisfies $\hat{\mathbf{t}}_\alpha^{i} \cdot \mathbf{e}_x = \cos\theta_\alpha^i$ and $\hat{\mathbf{t}}_\alpha^{i} \cdot \mathbf{e}_y = \sin\theta_\alpha^i$.  Then
\begin{equation}
    \begin{split}
	\mathsf{Q}_\alpha : \mathsf{E} &= \frac{\kappa}{L_\alpha} \sum_{i=0}^{Z-1} l^i_\alpha  \cos\theta^i_\alpha \sin\theta_\alpha^i  
	= \frac{\kappa}{2L_\alpha} \sum_{i=0}^{Z-1} l^i_\alpha \sin(2\theta_\alpha^i )\equiv  \kappa \mathsf{Q}_{\alpha, xy}.  
    \end{split}
\end{equation}
Similarly, noting that $\mathsf{E} \cdot \hat{\mathbf{t}}_\alpha^{i} = \kappa \sin\theta^i_\alpha \mathbf{e}_x$ and $\hat{\mathbf{t}}_\alpha^{i} \cdot \mathsf{E} \cdot \hat{\mathbf{t}}_\alpha^{i} = (\kappa/2)\sin(2\theta_\alpha^i)$, we have
\begin{equation}
	\begin{split}
		\mathsf{B}_\alpha : \mathsf{E} &= \frac{\kappa}{L_\alpha} \sum_{i=0}^{Z-1} l^i_\alpha \left[ \vphantom{\tfrac{1}{2}} (\cos\theta_\alpha^i \mathbf{e}_x + \sin\theta_\alpha^i \mathbf{e}_y)\sin\theta_\alpha^i \mathbf{e}_x 
		+ \sin\theta_\alpha^i \mathbf{e}_x(\cos\theta_\alpha^i\mathbf{e}_x + \sin\theta_\alpha^i \mathbf{e}_y) \right. \\
		& \qquad\qquad \left. - (\cos\theta_\alpha^i\mathbf{e}_x + \sin\theta_\alpha^i \mathbf{e}_y)(\cos\theta_\alpha^i\mathbf{e}_x 
        + \sin\theta_\alpha^i \mathbf{e}_y)\tfrac{1}{2}\sin(2\theta_\alpha^i)  \right] \\
		&= \frac{\kappa}{L_\alpha} \sum_{i=0}^{Z-1} l^i_\alpha \left[ \mathbf{e}_x\mathbf{e}_x \left(\tfrac{3}{4}\sin(2\theta_\alpha^i) - \tfrac{1}{8}\sin(4\theta_\alpha^i) \right) \right. 
		+ \mathbf{e}_y\mathbf{e}_y \left(\tfrac{1}{8}\sin(4\theta_\alpha^i) - \tfrac{1}{4}\sin(2\theta_\alpha^i) \right) 
		\\
		& \qquad \qquad \left. +  (\mathbf{e}_x\mathbf{e}_y + \mathbf{e}_y\mathbf{e}_x) \left(\tfrac{3}{8} + \tfrac{1}{8}\cos(4\theta_\alpha^i) - \tfrac{1}{2}\cos(2\theta_\alpha^i) \right) 
		\right].
	\end{split}
\end{equation}
Thus from \eqref{eq:tissue_pert_stress}, we have
\begin{equation}
	\Delta \boldsymbol{\sigma}^{\mathcal{M}}_{xy} = - \frac{1}{A^{\mathcal{M}}} \sum_\alpha^{N_c}  L_\alpha^2 \mathsf{Q}_\alpha : (\Gamma \mathsf{E}+\mu \dot{\mathsf{E}}) \mathsf{Q}_{\alpha, xy} + T_\alpha L_\alpha (\mathsf{B}_\alpha : \mathsf{E})_{xy}.
	\label{eq:shmd}
\end{equation}
To evaluate the elastic shear modulus of the disordered monolayer, we set $\Delta \sigma^{\mathcal{M}}_{xy}=-\kappa G_e$ with $\dot{\kappa}=0$.  The shear modulus is given by
\begin{equation}
	\label{eq:disordered_shearmod}
    \begin{split}
	G_e &=  \frac{1}{A^{\mathcal{M}}} \sum_\alpha^{N_c} \left[  \frac{\Gamma}{4} \left( \sum_{i=0}^{Z-1} l^i_\alpha \sin(2\theta_\alpha^i ) \right)^2 
    + T_\alpha \sum_{i=0}^{Z-1} l^i_\alpha \left(\tfrac{3}{8} + \tfrac{1}{8}\cos(4\theta_\alpha^i) - \tfrac{1}{2}\cos(2\theta_\alpha^i) \right) \right].
    \end{split}
\end{equation}
Equation (\ref{eq:disordered_shearmod}) recovers previous predictions for the shear modulus of periodic monolayers, where all cells are perfect hexagons ($L_{\alpha}^2 = 8 \sqrt{3} A_\alpha $, all edges have equal length, $\theta_\alpha^i = 2\pi i/6$ and the terms with sums over $\cos$ and $\sin$ vanish) \citep{nestorbergmann2017, murisic2015}; however it extends these results by allowing the direct evaluation of the shear modulus for a disordered monolayer.  Figure~\ref{fig:disordered_bulk_shear_mod}B demonstrates how $G_e$, as predicted by  \eqref{eq:disordered_shearmod}, varies across parameter space.  Interestingly, the tissue becomes less resistant to shear as it becomes increasingly resistant to isotropic deformations.  For comparison, Figure~\ref{fig:disordered_bulk_shear_mod}C shows the computationally simulated shear modulus, directly evaluated from the global perturbation stress tensor as $\Delta \sigma^{\mathcal{M}}_{12}=-\kappa G_e$, following a 1\% simple shear deformation ($\kappa=0.01$) on the simulated monolayers.  There is excellent agreement between the analytic and simulated results. 

For a sufficiently large disordered but isotropic monolayer, we might assume that the terms with sums over $\cos$ and $\sin$ in \eqref{eq:disordered_shearmod} vanish when summed over all cells (the degree of anisotropy can be assessed with \eqref{eq:tissue_isotropy}).  The elastic shear modulus for an isotropic monolayer is then approximated by
\begin{equation}
    \label{eq:disordered_shearmod_iso}
	G_e^{\mathrm{iso}} \approx \frac{3}{8A^{\mathcal{M}}} \sum_\alpha^{N_c} \Gamma L_\alpha(L_\alpha-L_0),
\end{equation}
showing how $G_e^{\mathrm{iso}}$ falls to zero as the tension in each cell approaches zero.  The percentage difference between $G_e$ and $G_e^{\mathrm{iso}}$ in example simulated isotropic tissues with 800 cells is shown in Figure~\ref{fig:disordered_bulk_shear_mod}D, showing close agreement ($<2\%$ relative difference) almost everywhere across region II.  Discrepancies arise only close to the region I/II boundary, where $G_e$ and $G_e^{\mathrm{iso}}$ both approach zero.  It is notable that the dynamic shear resistance term $\mu\dot{\mathsf{E}}$ in (\ref{eq:shmd}) is discarded under the approximation that leads to (\ref{eq:disordered_shearmod_iso}).  

\subsection{Composite isotropic and shear deformations}

\begin{figure} 
	\centering
	\includegraphics[width=0.55\textwidth]{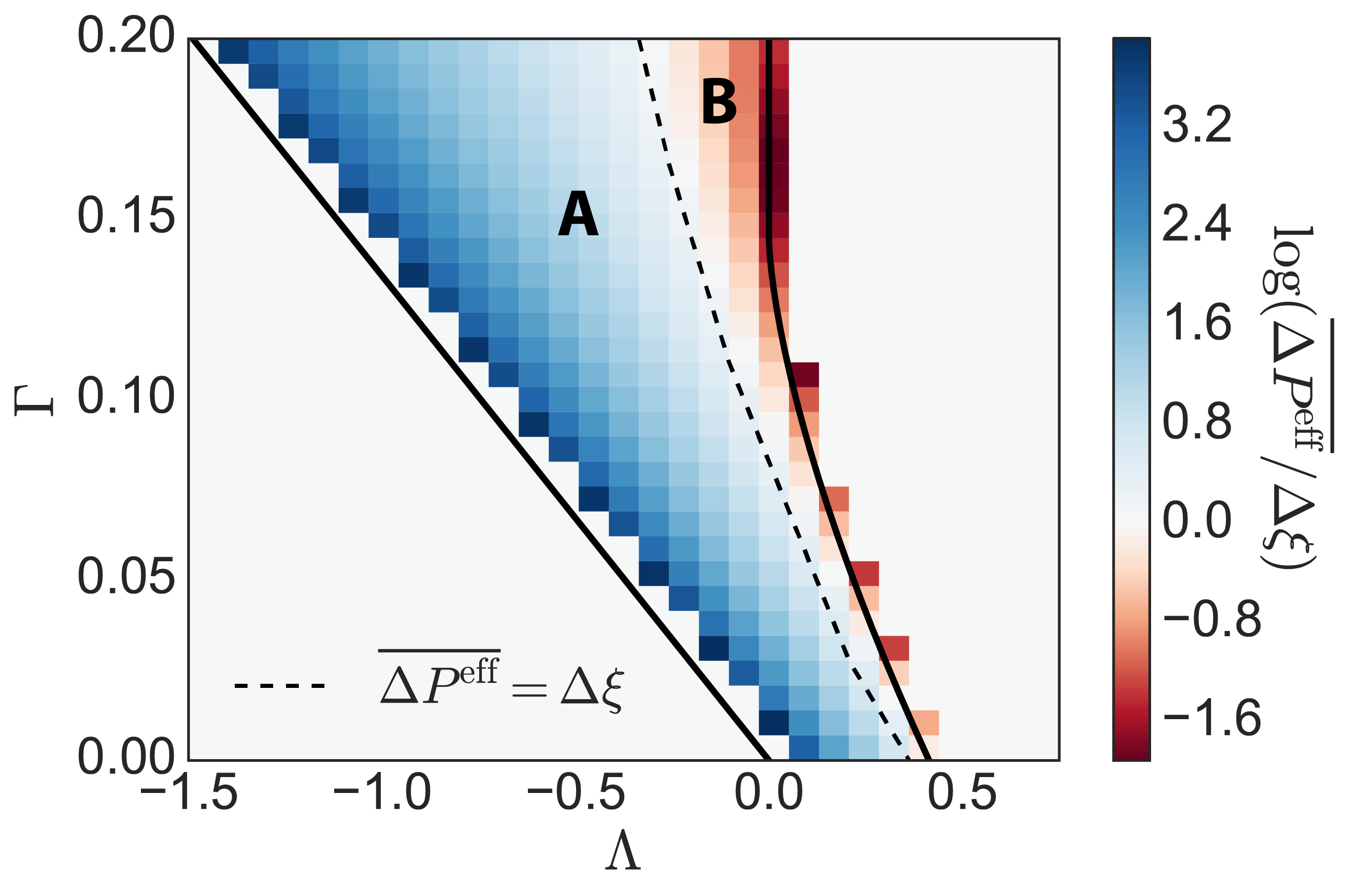}
	\caption{Heat map across discrete intervals of $(\Lambda,\Gamma)$ parameter space showing the value of $\log(\overline{\Delta{P}^{\mathrm{eff}}}/\hat{\xi})$.  The dashed line represents the contour where $\overline{\Delta{P}^{\mathrm{eff}}} = \hat{\xi}$: tissues show dominant resistance to area change in the shaded region to the left of the dashed line ({\color{red}where $\log(\overline{\Delta P^{\mathrm{eff}}}/\Delta\xi)>0$, A}), and to shear in the region to the right ({\color{red}where $\log(\overline{\Delta P^{\mathrm{eff}}}/\Delta\xi)<0$, B}).  A log scale is used to help display the differences across a large range of values. The monolayers used for all heat maps were the same as those used in Figure~\ref{fig:disordered_bulk_shear_mod}, where each datapoint is taken as an average from 5 realisations of a monolayer with 800 cells, with $\overline{P^\text{eff}}=0$. }
    \label{fig:stress_ratios}
\end{figure}

Finally, we recall that the perturbation stress tensor has eigenvalues $\Delta \sigma^{\mathcal{M}}_\pm = -\overline{\Delta{P}^{\mathrm{eff}}}\pm \Delta\xi$, where $(-\overline{\Delta{P}^{\mathrm{eff}}},- \overline{\Delta{P}^{\mathrm{eff}}})$ and $(\Delta{\xi},-\Delta\xi)$ are the eigenvalues of the isotropic and deviatoric (shear) components of \eqref{eq:tissue_pert_stress} respectively. 
Figure~\ref{fig:stress_ratios} shows how $\overline{\Delta{P}^{\mathrm{eff}}}/\Delta{\xi}$ varies across parameter space for isotropic monolayers subjected to static strain $\mathsf{E}^X$ (see \eqref{eq:composite_strain}) of amplitude 0.01, which induces both an isotropic and deviatoric stress response.  The parameter space partitions into a region that is more resistant to area change (region A) and one that is more resistant to shear (region B).  Figure~\ref{fig:stress_ratios} highlights how the isotropic stress, resisting area change, falls dramatically near the region II/III boundary.   

\section{Discussion} 

Previous reports have found that internal patterning in tissues can be linked to the mechanical properties of the material \citep{donev2005, Majmudar2005, honnell1990}.  We find that stretching induces ordering within the tissue, with cells being elongated on average in the direction of stretch, consistent with previous observations {\color{red}\citep{sugimura2013, wyatt2015,harris2012,xu2015}}.  Inferring relative stresses using the vertex-based model provides additional insight: distinguishing cells that are under net tension (with positive isotropic stress $P_\alpha^{\mathrm{eff}}>0$) from those under net compression, we find strong alignment of the former with the direction of stretch (Figure~2C) and of the latter with the perpendicular direction, in response to the imposed compressive stress, while retaining heterogeneity at the single-cell level.  This feature emerges strongly in direct simulations also (Figure~\ref{fig:example_stretch}).  Increased spatial organisation of cells is associated with anisotropic mechanical properties, which we characterised by deriving an explicit tissue-level stress/strain/strain-rate relationship (\ref{eq:dpef}, \ref{eq:dtj}, \ref{eq:tissue_pert_stress}) describing the response of a pre-stressed tissue to small-amplitude homogeneous deformations.  

The stress tensor we employed builds on the formulation derived by \citet{nestorbergmann2017} and others \citep{ishihara2012, guirao2015}, neglecting non-planarity \citep{hannezo2014, bielmeier2016}, curved cell edges \citep{brodland2014, ishimoto2014} and further refinements but including internal dissipation due to dynamic area and length changes of individual cells in a way that naturally complements the assumed strain energy.  Our formulation ensures no net change in internal energy under a homogeneous deformation (Appendix~\ref{sec:mappings_derivation}) {\color{red}and is suited to describing the viscoelastic properties of freely suspended monolayers, as described by \cite{harris2012}.}  The model is in the spirit of, but differs from, that of Okuda \textit{et al.} \cite{okuda2015}, who proposed a drag force depending on an average of nearby vertex velocities.    The linearized stress/strain relationship (\ref{eq:dpef}, \ref{eq:dtj}, \ref{eq:tissue_pert_stress}) does not include additional dissipative effects of substrate drag or irreversible cell rearrangements.  A framework for including additional plastic stresses and strains has been proposed within a coarse-grained model \cite{ishihara2017cells}; simulations of large-amplitude plastic tissue deformations under external loading using a discrete cell model are provided by \cite{pathmanathan2009computational}.

Under the present vertex model, the perturbation stress of a pre-stressed tissue is given by the area-weighted sum of perturbation stresses of the individual cells \eqref{eq:tissue_pert_stress}.  This leads to an expression for the fourth-order stiffness tensor $\mathsf{C}$ (\ref{eq:cd}a), which describes how anisotropic tissues resist deformation through reversible elastic deformations, {\color{red}and its viscous analogue $\mathsf{D}$ (\ref{eq:cd}b)}.  This formulation extends previous approaches to upscaling the vertex-based model in spatially periodic \citep{murisic2015} and disordered isotropic networks \citep{nestorbergmann2017}.  Exact expressions for elastic moduli for a given monolayer realisation are provided as explicit sums (for isotropic monolayers, these are (\ref{eq:bulk}) and (\ref{eq:disordered_shearmod})).  Further work is required to derive \textit{a priori} predictions for the behaviour of these quantities over multiple monolayer realisations.  However a crude simplification of the estimate of the elastic shear modulus (\ref{eq:disordered_shearmod_iso}) for a disordered isotropic monolayer is accurate across the bulk of region II of  parameter space, but not close to the phase transition along the region I/II boundary where the shear modulus approaches zero.  Our predictions can be compared with those of Merzouki \textit{et al.}~\cite{merzouki2016}, who used simulations to impose stress and measure strain of a periodic hexagonal monolayer in order to infer bulk elastic parameters.  Our results are broadly consistent with theirs, including evidence of non-monotonic behaviour close to the region II/III boundary  (revealed most clearly in the stress ratios plotted in Figure~\ref{fig:stress_ratios}).  {\color{red}Likewise Xu et al.~\cite{xu2015} report reduced stress under uniaxial strain for increasing $\Gamma$ and $\Lambda$, mirroring the bulk elastic response in Figure~\ref{fig:disordered_bulk_shear_mod}D.}  By comparing the relative size of isotropic and shear stresses induced by a strictly uniaxial deformation, the vertex model can also be used to partition parameter space into regions of higher shear modulus and lower bulk modulus, and \textit{vice versa} (Figure~\ref{fig:stress_ratios}).  This may be relevant to phases during development when tissues undergo extreme shape changes and may have a bearing on the different mechanical environments of of epithelial tissues in various organ systems.  



In summary, this study demonstrates how loading organises the cell-scale stress field in a stretched monolayer and how mechanical viscoelastic moduli of disordered or anisotropic cellular monolayers can be determined as explicit sums over cells.  Further steps towards deriving well-grounded homogenized descriptions of such media will require assessment of the statistical distributions of different cell classes over the plane.

\begin{acknowledgments}
ANB was supported by a BBSRC studentship and EJ by a Wellcome Trust studentship.  OEJ acknowledges EPSRC grant EP/K037145/1. SW was supported by a Wellcome Trust/Royal Society Sir Henry Dale Fellowship [098390/Z/12/Z].
\end{acknowledgments}

\appendix

\section{A homogeneous small-amplitude strain}
\label{sec:mappings_derivation}

We derive the mappings of key geometric quantities under a small-amplitude homogenous and symmetric strain, $\mathsf{E}$, which transforms position vectors as $\mathbf{R} \rightarrow \mathbf{R} + \mathsf{E}\cdot\mathbf{R}$, and then apply these mappings to the mechanical energy $U$.

We assume that all quantities, $X$, follow a mapping of the form $X \rightarrow X + \Delta X$, defined relative to the same cell and therefore temporarily drop the $\alpha$ subscript.  Tangents are defined as $\mathbf{t}^i = \mathbf{R}^{i+1} - \mathbf{R}^i$, giving
$\mathbf{t}^i + \Delta \mathbf{t}^i = \mathbf{R}^{i+1} +\mathsf{E} \cdot \mathbf{R}^{i+1} - \mathbf{R}^i - \mathsf{E} \cdot \mathbf{R}^i$,
and hence $\Delta \mathbf{t}^i = \mathsf{E} \cdot \mathbf{t}^i$.
 The length of an edge is given by $l^i = ( \mathbf{t}^i \cdot \mathbf{t}^i )^{\tfrac{1}{2}}$. To linear order,
\begin{equation}
	\label{eq:length_mapping}
    \begin{split}
	l^i + \Delta l^i &= \left[ (\mathbf{t}^i + \Delta \mathbf{t}^i) \cdot (\mathbf{t}^i + \Delta \mathbf{t}^i) \right]^{\tfrac{1}{2}} 
	\approx \left[ (l^i)^2 + 2 \mathbf{t}^i \cdot \mathsf{E} \cdot \mathbf{t}^i \right]^{\tfrac{1}{2}} 
	\approx l^i ( 1 + \hat{\mathbf{t}}^i \cdot \mathsf{E} \cdot \hat{\mathbf{t}}^i ),
    \end{split}
\end{equation}
demonstrating that $\Delta l^i = l^i (\hat{\mathbf{t}}^i \cdot \mathsf{E} \cdot \hat{\mathbf{t}}^i) $.
The cell perimeter is $L = \sum_{i=0}^{Z-1} l^i $, so that
\begin{equation}
	\label{eq:Q_pert}
	\begin{split}
	L + \Delta L &= \sum_{i=0}^{Z-1} l^i ( 1 + \hat{\mathbf{t}}^i \cdot \mathsf{E} \cdot \hat{\mathbf{t}}^i ) 
	= L + \sum_{i=0}^{Z-1} l^i ( \hat{\mathbf{t}}^i \cdot \mathsf{E} \cdot \hat{\mathbf{t}}^i ) 
	= L(1 + \mathsf{Q}:\mathsf{E}).
	\end{split}
\end{equation}
Thus $\Delta L = L \mathsf{Q}:\mathsf{E}$. 
It follows that 
$	T + \Delta T = \Gamma(L + \Delta L - L_0) 
	= T + \Gamma L \mathsf{Q}:\mathsf{E}$.
Similarly, $l^i \hat{\mathbf{t}}^i \rightarrow l^i \hat{\mathbf{t}}^i + \Delta (l^i \hat{\mathbf{t}}^i)$. From the product rule,
$	\Delta (l^i \hat{\mathbf{t}}^i) = (\Delta l^i)\hat{\mathbf{t}}^i + l^i (\Delta \hat{\mathbf{t}}^i)$.
Now, $\Delta (l^i \hat{\mathbf{t}}^i) = l^i \mathsf{E}\cdot \hat{\mathbf{t}}^i$ and using \eqref{eq:length_mapping} gives
$l^i \mathsf{E}\cdot \hat{\mathbf{t}}^i = l^i (\hat{\mathbf{t}}^i \cdot \mathsf{E} \cdot \hat{\mathbf{t}}^i) \hat{\mathbf{t}}^i  + l^i (\Delta \hat{\mathbf{t}}^i)$.
Thus, $\Delta \hat{\mathbf{t}}^i = \mathsf{E}\cdot \hat{\mathbf{t}}^i - (\hat{\mathbf{t}}^i \cdot \mathsf{E} \cdot \hat{\mathbf{t}}^i)\hat{\mathbf{t}}^i $.  Note that $\hat{\mathbf{t}}^i \cdot (\Delta \hat{\mathbf{t}}^i) = 0$, ensuring that unit vectors are rotated but not stretched.


We now consider the deformation $L\mathsf{Q} \rightarrow L\mathsf{Q} + \Delta(L\mathsf{Q})$, writing
\begin{equation}
	\begin{split}
	L\mathsf{Q} + \Delta(L\mathsf{Q}) &= \sum_{i=0}^{Z-1} [l^i \hat{\mathbf{t}}^{i} + \Delta (l^i \hat{\mathbf{t}}^{i})][ \hat{\mathbf{t}}^{i} + \Delta \hat{\mathbf{t}}^{i} ] \\
	&= \sum_{i=0}^{Z-1}  [l^i \hat{\mathbf{t}}^{i} + l^i \mathsf{E}\cdot \hat{\mathbf{t}}^i ][ \hat{\mathbf{t}}^{i} + \mathsf{E}\cdot \hat{\mathbf{t}}^i 
    - (\hat{\mathbf{t}}^i \cdot \mathsf{E} \cdot \hat{\mathbf{t}}^i)\hat{\mathbf{t}}^i ] \\
	&\approx \sum_{i=0}^{Z-1} l^i \hat{\mathbf{t}}^{i} \hat{\mathbf{t}}^{i} + \sum_{i=0}^{Z-1} \left\{ l^i \hat{\mathbf{t}}^{i}[\mathsf{E} \cdot \hat{\mathbf{t}}^{i} 
    - (\hat{\mathbf{t}}^{i}\cdot \mathsf{E} \cdot \hat{\mathbf{t}}^{i}) \hat{\mathbf{t}}^{i} ] + l^i (\mathsf{E} \cdot \hat{\mathbf{t}}^{i}) \hat{\mathbf{t}}^{i} \right\}
	\end{split}
\end{equation}
to linear order.  Thus we see that 
	$\Delta(L\mathsf{Q}) = L\mathsf{B}:\mathsf{E}$,
where $\mathsf{B}$ in component form is
{\color{red}
\begin{equation}
 \left\{ \mathsf{B}_{\alpha}\right\}_{pqrs} = \frac{1}{L_\alpha} \sum_{i=0}^{Z_\alpha-1} l^i_\alpha \left[ 
\tfrac{1}{2} \left(    \hat{{t}}_{\alpha,p}^{i} \mathsf{I}_{qr}  \hat{{t}}_{\alpha,s}^{i} +
\hat{{t}}_{\alpha,q}^{i} \mathsf{I}_{pr}  \hat{{t}}_{\alpha,s}^{i} +
\hat{{t}}_{\alpha,p}^{i} \mathsf{I}_{qs}  \hat{{t}}_{\alpha,r}^{i} +
\hat{{t}}_{\alpha,q}^{i} \mathsf{I}_{ps}  \hat{{t}}_{\alpha,r}^{i}  \right)
    - \hat{{t}}_{\alpha,p}^{i} \hat{{t}}_{\alpha,q}^{i} \hat{{t}}_{\alpha,r}^{i} \hat{{t}}_{\alpha,s}^{i} \right]
  \label{eq:B}
\end{equation}
ensuring that $\{\mathsf{B}_\alpha\}_{pqrs}=\{\mathsf{B}_{\alpha}\}_{qprs}=\{\mathsf{B}_{\alpha}\}_{pqsr}$.}  




\begin{figure} 
	\centering
	\includegraphics[width=0.3\textwidth]{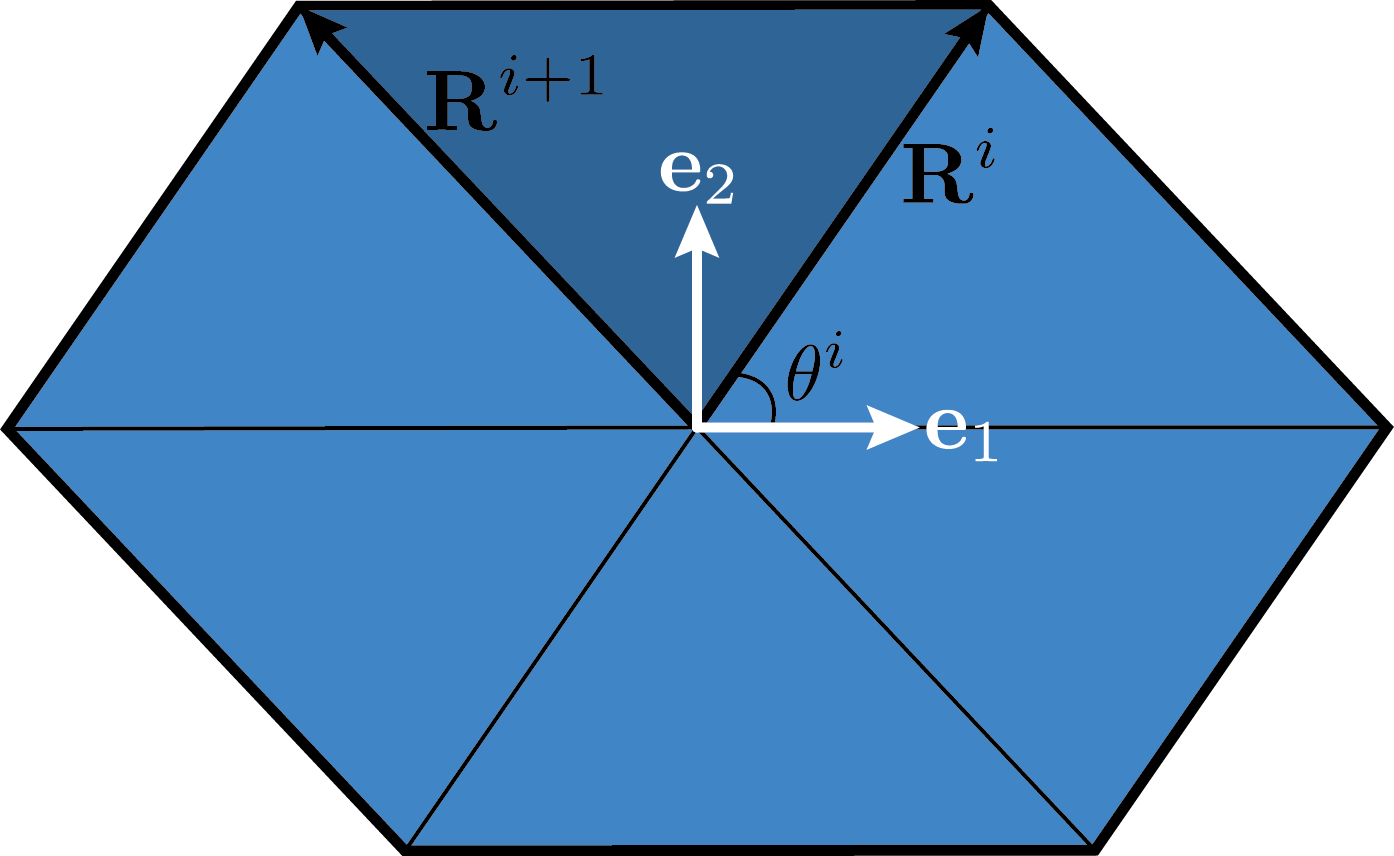}
	\caption{Representative geometry used to calculate the area of a cell.  }
    \label{fig:triangle_geometry}
\end{figure}

To evaluate area changes, we begin by considering the area of a sub-triangle, $\bigtriangleup = \{ \mathbf{R}_\alpha, \mathbf{R}^i_{\alpha}, \mathbf{R}^{i+1}_\alpha \}$, comprised of the cell centroid and two vertices from an edge (Figure~\ref{fig:triangle_geometry}). We have used the $\alpha$ subscript notation for clarity in defining the centroid, but drop it again from this point.  The area of this triangle is 
	$A^\bigtriangleup = \frac{1}{2} \hat{\mathbf{z}} \cdot ( \mathbf{R}^i \times \mathbf{R}^{i+1} )$.
Applying $\mathsf{E}$ we have
\begin{equation}
	\label{eq:a_map_first}
	\begin{split}
	A^\bigtriangleup + \Delta A^\bigtriangleup &= \tfrac{1}{2} \hat{\mathbf{z}} \cdot ( \mathbf{R}^i + \mathsf{E} \cdot \mathbf{R}^i) \times (\mathbf{R}^{i+1} + \mathsf{E} \cdot \mathbf{R}^{i+1} ) \\
	&\approx \tfrac{1}{2} \hat{\mathbf{z}} \cdot ( \mathbf{R}^i \times \mathbf{R}^{i+1} + \mathbf{R}^i \times \mathsf{E} \cdot \mathbf{R}^{i+1}  
	+ \mathsf{E} \cdot \mathbf{R}^i \times \mathbf{R}^{i+1} )
	\end{split}
\end{equation}
to linear order.  Since $\mathsf{E}$ is symmetric, we can make use of the spectral theorem to write $\mathsf{E} = \lambda_1 \mathbf{e}_1 \mathbf{e}_1 + \lambda_2 \mathbf{e}_2 \mathbf{e}_2$, where $(\mathbf{e}_1, \mathbf{e}_2)$ form an orthonormal basis of eigenvectors.  Inserting into \eqref{eq:a_map_first} we have
\begin{equation}
    \begin{split}
\Delta A^\bigtriangleup &= \tfrac{1}{2} \hat{\mathbf{z}} \cdot [ \lambda_1 \mathbf{R}^i \times \mathbf{e}_1 \mathbf{e}_1 \cdot \mathbf{R}^{i+1} 
     + \lambda_2 \mathbf{R}^i \times \mathbf{e}_2 \mathbf{e}_2 \cdot \mathbf{R}^{i+1} \\
 & \qquad \qquad   + \lambda_1  (\mathbf{e}_1 \mathbf{e}_1\cdot \mathbf{R}^i) \times \mathbf{R}^{i+1} 
    + \lambda_2( \mathbf{e}_2 \mathbf{e}_2 \cdot\mathbf{R}^i )\times \mathbf{R}^{i+1} ].
    \end{split}
\end{equation}
Defining $\theta^i$ and $\theta^{i+1}$ such that $\lambda_1 \mathbf{R}^i \times \mathbf{e}_1 = - \lambda_1 |\mathbf{R}^i| \sin \theta^i \hat{\mathbf{z}}$,  and $\mathbf{e}_1 \cdot \mathbf{R}^{i+1} = |\mathbf{R}^{i+1}| \cos \theta^{i+1}$ (see Figure~\ref{fig:triangle_geometry}) gives
\begin{equation}
	\begin{split}
	\Delta A^\bigtriangleup &= \tfrac{1}{2} |\mathbf{R}^{i}| |\mathbf{R}^{i+1}| [ -\lambda_1 \sin\theta^i \cos\theta^{i+1} 
    + \lambda_2 \sin(\tfrac{\pi}{2} - \theta^{i}) \cos(\tfrac{\pi}{2} - \theta^{i+1}) \\ 
    & \qquad\qquad\qquad\qquad + \lambda_1 \cos\theta^i \sin\theta^{i+1} 
    - \lambda_2 \cos(\tfrac{\pi}{2} - \theta^{i}) \sin(\tfrac{\pi}{2} - \theta^{i+1}) ] \\
	&= \tfrac{1}{2} |\mathbf{R}^{i}| |\mathbf{R}^{i+1}| [ \lambda_1 \sin(\theta^{i+1} - \theta^{i} ) + \lambda_2 \sin(\theta^{i+1} - \theta^{i}) ] 
	= \mathrm{Tr}(\mathsf{E}) A^\bigtriangleup.
	\end{split}
\end{equation}
Summing over sub-triangles gives 
	$\Delta A = \mathrm{Tr}(\mathsf{E}) A$.

Having shown that under the deformation $\mathbf{R} \rightarrow \mathbf{R} + \mathsf{E}\cdot\mathbf{R}$, lengths and areas of individual cells transform according to $L_\alpha \rightarrow L_\alpha(1 + \mathsf{Q}_\alpha : \mathsf{E} )$, $A_\alpha \rightarrow A_\alpha(1 + \mathrm{Tr}(\mathsf{E}))$, it is straightforward to determine the associated change in recoverable mechanical energy (\ref{eq:final_nondim}) as
\begin{equation}
\Delta U=\sum_\alpha \Delta U_\alpha=\sum_\alpha P_\alpha  \Delta A_\alpha + T_\alpha \Delta L_\alpha = \sum_\alpha (P_\alpha A_\alpha \mathsf{I}+T_\alpha L_\alpha \mathsf{Q}_\alpha) : \mathsf{E}= \boldsymbol{\Sigma}^{(e)}: \mathsf{E},
\end{equation}
where $\boldsymbol{\Sigma}^{(e)}\equiv -\sum A_\alpha \boldsymbol{\sigma}_\alpha^{(e)}$.  Here we have decomposed the stress in (\ref{eq:stres1}) into its elastic and viscous components $\boldsymbol{\sigma}_\alpha=\boldsymbol{\sigma}_\alpha^{(e)}+\boldsymbol{\sigma}_\alpha^{(v)}$.  The sign change between $\boldsymbol{\sigma}$ and $\boldsymbol{\Sigma}$ arises from the difference between the stresses exerted on, or by, a cell or tissue.  Similarly defining $\boldsymbol{\Sigma}^{(v)} \equiv - \sum A_\alpha \boldsymbol{\sigma}_\alpha^{(v)}$, and
noting that for this small deformation $\dot{A}_\alpha=A_\alpha \mathrm{Tr}(\dot{\mathsf{E}})$, it follows that 
\begin{equation}
\boldsymbol{\Sigma}^{(v)}: \dot{\mathsf{E}} =\sum_\alpha \gamma A_\alpha \dot{A}_\alpha \mathrm{Tr}(\dot{\mathsf{E}})+\mu L_\alpha \dot{L}_\alpha \mathsf{Q}_\alpha:\dot{\mathsf{E}} = \sum \gamma \dot{A}_\alpha^2+\mu \dot{L}_\alpha^2 =  \Phi,
\end{equation} 
where $\Phi=\sum_\alpha \Phi_\alpha$ is the dissipation rate (\ref{eq:diss}).  In the absence of neighbour exchanges, which would contribute additional stresses and deformations (treated in a coarse-grained approximation by Ishihara \textit{et al.} \cite{ishihara2017}), the total rate of change of internal energy of the system is therefore $\dot{\mathcal{E}}=\boldsymbol{\Sigma}:\dot{\mathsf{E}}=(\boldsymbol{\Sigma}^{(e)}+\boldsymbol{\Sigma}^{(v)}):\dot{\mathsf{E}}=\dot{U}+\Phi=0$ (by (\ref{eq:cons})).   Positing the thermodynamic relation $\Delta \mathcal{E}=T\Delta S + \Delta U$, where $\Delta S$ is an entropy change at temperature $T$, it follows that for the imposed deformation $\mathsf{E}$, $T\dot{S}=-\dot{U}=\Phi\geq 0$.

\section{Experimental methods}
\label{sec:appexp}

\textit{Xenopus laevis} female frogs were pre-primed 4-7~days in advance with 50~units of Pregnant Mare's Serum Gonadotrophin (Intervet UK) and then primed with 500~units of Human Chorionic Gonadotrophin (Intervet UK) 18~hr before use as detailed in \cite{woolner2008myosin}.  Each frog was housed individually overnight and transferred to room temperature 1x Marc's Modified Ringers solution (100 mM NaCl, 2~mM KCl, 1~mM MgCl, and 5~mM HEPES [pH 7.4]) at least 2~hr prior to egg collection.  \textit{In vitro} fertilisation was performed as described previously \cite{woolner2008myosin} and embryos were dejellied using 2\% cysteine (in 0.1x~MMR, pH 7.8-8.0). Embryos were microinjected with a needle volume of 5 and 2.5~nl at the two- or four-cell stages respectively into all cells, using a Picospritzer III (Parker instrumentation) with embryos submerged in 0.1x~MMR plus 5\% Ficoll. RNA was synthesised as described previously \cite{sokac2003cdc42} and microinjected at the following needle concentrations: 0.5~mg/ml GFP-$\alpha$-tubulin; 0.1~mg/ml cherry-histone2B \cite{kanda1998histone}.

Animal cap tissue was dissected from the embryo at stage 10 of development (early gastrula stage) following a protocol previously described by Joshi and Davidson (2010)~\cite{joshi2010live}, and cultured on a 20~mm x 20~mm x 1~mm elastomeric PDMS membrane coated with fibronectin (incubated at 4$^{\circ}$C overnight with 1~ml of 10~$\mu$g/ml fibronectin). The fibronectin was removed and each membrane was subsequently washed 3 times with 1x PBS followed by 2 washes with Danilchik's for Amy explant culture media (DFA; 53~mM NaCl2, 5~mM Na2CO3, 4.5~mM potassium gluconate, 32~mM sodium gluconate, 1~mM CaCl2, 1~mM MgSO4) prior to the introduction of each animal cap. Animal cap explants were excised using forceps and hair-knives to make neat squares of tissue. Each explant was transferred to a PDMS membrane filled with DFA and a coverslip with vacuum grease at each end was placed over the top to ensure the explant adhered to the fibronectin-coated membrane. Each membrane was then incubated at 18$^{\circ}$C for at least 2~hr prior to imaging.

Each PDMS membrane was attached to a stretch apparatus (custom made by Deben UK Limited) fixed securely to the stage of a Leica TCS SP5 AOBS upright confocal and a 0.5~mm or 8.6~mm uniaxial stretch was applied for control (unstretched) and stretched samples respectively. 
Images were collected on the Leica TCS SP5 AOBS upright confocal using a 20x/0.50 HCX Apo U-V-I (W (Dipping Lens)) objective and 2x (or 1x) confocal zoom. The confocal settings were as follows, pinhole 1 Airy unit, scan speed 400~Hz bidirectional, format 512 x 512 (or 1024 x 1024). Images were collected using hybrid detectors with the following detection mirror settings; HyD2 FITC 500-550~nm; HyD4 Texas red 590-690~nm using the 488~nm (20\%) and 543~nm (100\%) laser lines respectively. Images were collected sequentially. The distance between each optical stack was maintained at 4.99~$\mu$m and the time interval between each capture was restricted to 20~s, with each samples being imaged for up to 2.5~hr. Only the maximum intensity projections of these 3D stacks are shown in the results.

\begin{figure}
\includegraphics[width=0.6\textwidth]{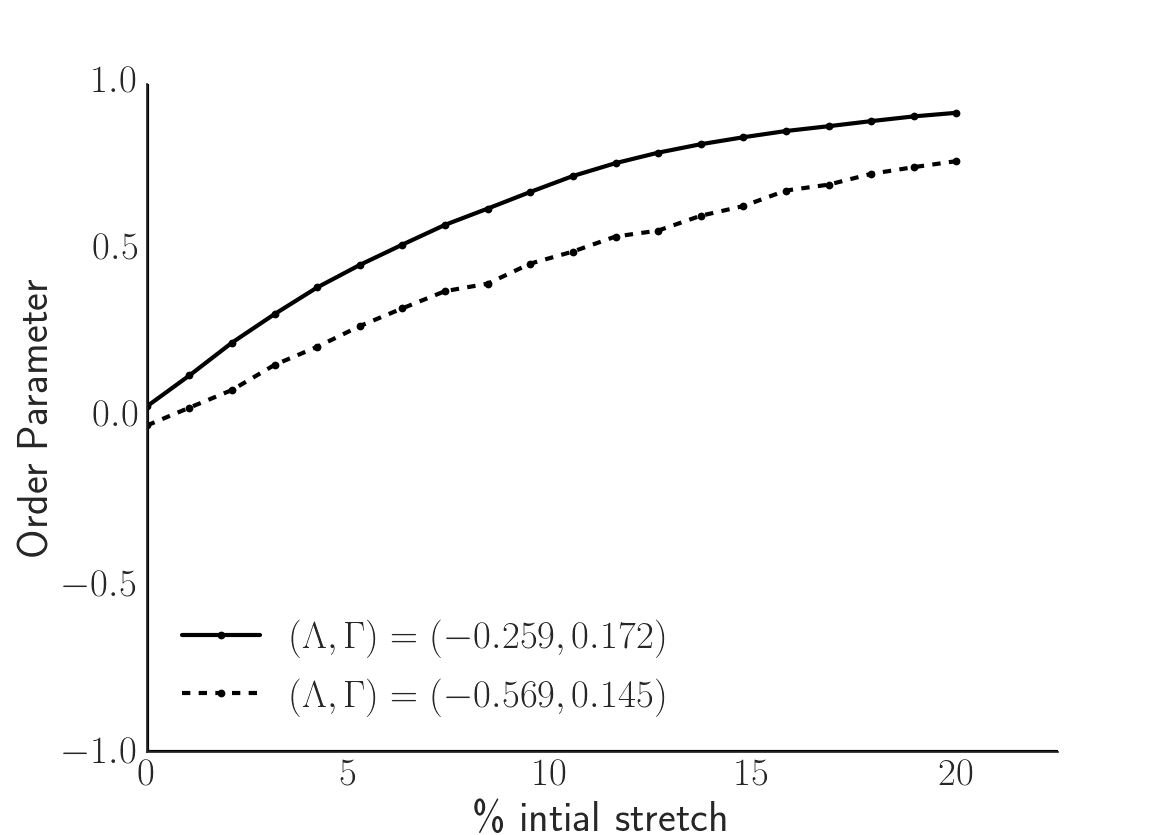}
\caption{The order parameter $Q$, for the realisations plotted in Figure~\ref{fig:stress_vs_stretch}B.}
\label{fig:op}
\end{figure}

{\color{red}
\section{Order parameter}
\label{app:c}

A more traditional measure of spatial disorder is provided by the order parameter $Q=\langle \cos 2\theta \rangle$ where $\theta$ measures the angle of the principal axis of the shape tensor of each cell ($\sum_{i=0}^{Z_\alpha-1} \mathbf{R}_{\alpha}^i \otimes \mathbf{R}_\alpha^i$) with respect to the direction of stretch and the average is taken over all the cells in the monolayer.   Figure~\ref{fig:op} illustrates the evolution of $Q$ for the stretch realisations illustrated in Figure~\ref{fig:stress_vs_stretch}B.  It is notable that while the two examples are quite similar by this geometric measure, the shear stress (Figure~\ref{fig:stress_vs_stretch}B) is substantially lower for parameters closer to the region I/II boundary in parameter space (Figure~\ref{fig:geometry}B).}

\bibliography{abbreviated}

\end{document}